\documentclass[onecolumn,sort&compress]{elsarticle}
\usepackage{ifpdf}
\usepackage[usenames]{color}
\usepackage{bm}
\usepackage{times}
\usepackage{graphicx}
\usepackage{amssymb}
\usepackage{amsmath}
\usepackage{simplewick}
\usepackage{hyperref}
\usepackage{color}
\usepackage{float}
\usepackage{caption}
\usepackage{rotating}

\usepackage{tikz}
\usetikzlibrary{decorations.pathmorphing}
\usetikzlibrary{decorations.markings}
\usetikzlibrary{shapes,arrows}
\usetikzlibrary{arrows.meta}
\tikzset{middlearrow/.style={
        decoration={markings,
           mark= at position 0.5 with {\arrow{#1}} ,
        },
        postaction={decorate}
   }
}

\journal{Computer Physics Communications}
\hypersetup{
      colorlinks=true,
      citecolor=blue,
      linkcolor=blue,
      urlcolor=blue}

\begin{document}

\begin{frontmatter}

\title{FACt: FORTRAN toolbox for calculating fluctuations in atomic condensates}

\author[a,b]{Arko Roy}
\author[a,f]{Sukla Pal \corref{author}}
\cortext[author] {Corresponding author.\\\textit{E-mail address:} 
        sukla.pal@otago.ac.nz}
\author[c]{S. Gautam }
\author[a]{D. Angom }
\author[d,e]{P. Muruganandam }

\address[a]{Physical Research Laboratory,
            Navarangpura, Ahmedabad 380 009, Gujarat, India
            }
\address[b]{Max-Planck-Institut f{\"u}r Physik komplexer Systeme, 
N{\"o}thnitzer Stra\ss e 38, 01187 Dresden, Germany
            }
\address[c]{Department of Physics,
            Indian Institute of Technology Ropar, 
            Rupnagar, Punjab 14001, India
           } 
\address[d]{Department of Physics,
            Bharathidasan University, Tiruchirappalli 620 024,India           
           }
\address[e]{Department of Medical Physics,
            Bharathidasan University, Tiruchirappalli 620 024,India
           }
\address[f]{Dodd-Walls Centre for Photonic and Quantum Technologies and Department of Physics, University of Otago, 
            Dunedin 9016, New Zealand}
\begin{abstract} 
 We develop a FORTRAN code to compute fluctuations in atomic condensates (FACt)
by solving the Bogoliubov-de Gennes (BdG) equations for two component 
Bose-Einstein condensate (TBEC) in quasi-two dimensions. The BdG equations are 
recast as matrix equations and solved self consistently. The code is suitable 
for handling quantum fluctuations as well as thermal fluctuations at 
temperatures below the critical point of  Bose-Einstein condensation. The 
code is versatile, and the ground state density profile and low energy 
excitation modes obtained from the code can be easily adapted to compute
different properties of TBECs --- ground state energy, overlap integral, 
quasi particle amplitudes of BdG spectrum, dispersion relation and structure 
factor and other related experimental observables.
\end{abstract}


\end{frontmatter}

{\bf PROGRAM SUMMARY}

\noindent
{\em Program Title:} FACt                                 \\
{\em Journal Reference:}                                      \\
{\em Catalogue identifier:}                                   \\
{\em Licensing provisions:} none                                   \\
{\em Programming language:}FORTRAN 90                         \\
{\em Computer:} Intel Xeon,                                   \\
{\em Operating system:} General                               \\
{\em RAM:} at least 1.5Gbytes per core.                        \\
{\em Number of processors used:} 1                              \\
{\em Supplementary material:} none                                 \\
{\em Classification:}                                         \\
{\em External routines/libraries:} ARPACK                        \\
{\em Subprograms used:}                                       \\
{\em Journal reference of previous version:}*                  \\
{\em Nature of problem:} Compute the ground state density profile, ground 
                         state energy and chemical potential for individual
                         species, evaluate the quasiparticle mode energies and 
                         corresponding amplitudes which can capture the
                         transformation of the modes against the change of the
                         parameters (intraspecies interaction, interspecies
                         interaction, anisotropy parameter etc.) using
                         Hartree-Fock Bogoliubov theory with the Popov
                         approximation. Calculate the overlap integral,
                         dispersion relation and structure factor.
 \\
{\em Solution method:}   In the first step, the pair of coupled
                         Gross-Pitaevskii equations (CGPEs)
                         are solved using split time-step 
                         Fourier pseudospectral method to 
                         compute the condensate density. To solve the BdG
                         equations,
                         as a basic input the first $N_b$ harmonic oscillator 
                         eigenstates are chosen as a basis to generate the BdG
                         matrix with dimension of
                         $4(N_b + 1)\times 4(N_b + 1)$.
                         Since the matrix size rapidly increases with $N_b$, 
                         \textsc{Arpack} routines are used to diagonalise the 
                         BdG matrix efficiently. To compute the fluctuation and
                         non-condensate density, a set of the low energy 
                         quasiparticle amplitudes above a threshold value of 
                         the Bose factor are considered. The equations are then 
                         solved iteratively till the condensate, and 
                         non-condensate densities converge to predefined 
                         accuracies. To accelerate the convergence we use the
                         method of successive under-relaxation (SUR). 
                     
{\em Restrictions:}

      For a large system size, if the harmonic oscillator basis size is also
      taken to be large, the dimension of the BdG matrix becomes huge. It may
      take several days to compute the low energy modes at finite temperature 
      and this package may be computationally expensive.



{\em Additional comments:} 

       After successful computation of this package, one should obtain the 
       equilibrium density profiles for TBEC, low energy 
       Bogoliubov modes and the corresponding quasiparticle amplitudes. In 
       addition, one can calculate the dispersion relation, structure factor,
       overlap integral, correlation function, etc. using this package with 
       minimal modifications. In the theory section of the manuscript, we have
       provided the expressions to compute the above quantities numerically.

{\em Running time: } 

      $\sim10$ minutes for the sample case. For self consistent 
      calculation with 15 iterations, it could take approximately 2 days for 
      the parameters specified in the manuscript.


\section{Introduction}
\label{intro}

The self-consistent Hartree-Fock-Bogoliubov theory with the Popov (HFB-Popov) 
approximation is an effective model to examine the fluctuations of equilibrium 
state solutions of trapped BEC at zero temperature as well as finite 
temperatures. The theory is in particular well suited to examine the 
evolution of the low-lying modes as a function of the interaction parameters, 
temperature or trapping parameters. It has been used extensively in 
single-species BEC to study finite temperature effects and mode 
energies \cite{griffin_96, hutchinson_97, dodd_98, gies_04}, and the
results are in good agreement with experimental results \cite{jin_97} at
low temperatures. The detailed and systematic information about the 
quasiparticle spectrum, both of single and multispecies condensate,
are described by the HFB formalism. In two-species BECs (TBECs), where the
phenomenon of phase-separation is important \cite{gautam_11, gautam_10},
the HFB-Popov approximation has been used in the miscible \cite{pires_08} and 
immiscible domain \cite{roy_14,roy_15, roy_16} to compute the low-lying modes. 

In the present work we report the development of a FORTRAN code which
implements the HFB-Popov theory to compute the low energy elementary 
excitations of the TBECs. At $T = 0$K, where 
only the quantum fluctuations are present in the system, the code captures the
essence of quantum fluctuations. These are important in the stabilization
of quantum droplets in binary 
BEC mixtures~\cite{cabrera_18, petrov_15, petrov_16, semeghini_18}. In our 
recent works~\cite{pal_17, pal_18} we have investigated the elementary 
excitations in radially symmetric and anisotropic TBECs using the present 
version of FACt. However, the main strength of HFB-Popov approximation is in 
encapsulating properties of trapped BEC at finite temperatures, which is more 
realistic and experimentally relevant. It must be emphasized that our code 
provides high precision and converged results for $T\ll T_c$ and computes the 
low energy excitation modes for TBECs in quasi two dimension. 
It is worth pointing out here that in 3D the fluctuations are 
less prominent and mean field theories work very well. That is not the case 
in 2D. The presence of fluctuations, both thermal and quantum, inhibits
real condensation in 2D because of Mermin-Wagner-Hohenberg (MWH) 
theorem~\cite{hohenberg_65,mermin_66};
but undergoes a Berezinskii-Kosterlitz-Thouless (BKT) phase 
transition~\cite{berezinskii_72,kosterlitz_72,kosterlitz_73,kosterlitz_16}.
With regard to the experimental realization of the box trapping
potential~\cite{ville_18}, our codes are timely and ideal to study 
2D multicomponent systems.
It must also be 
mentioned that the HFB-Popov has been used to study quantum and thermal 
fluctuations in optical lattices \cite{suthar_15, suthar_16}. 
It is to be mentioned here that the HFB-Popov equations belong
to the general class of linear response problems and very efficient 
numerical methods have been developed to solve these 
equations~\cite{bai_12,bai_14}.

An important feature of our implementation, which optimizes the computational 
requirements, is the absence of any constraints on the symmetry. That is, we 
implement the code in Cartesian coordinates. The basic and important advantage 
of this approach is that, our code is very general and applicable to the
anisotropic cases where the frequency of the trap in $x$ and $y$ directions 
are different.


\section{Finite temperature theory for two component BEC}
\label {fnite-temp-th}

 In the dilute limit, when the interparticle interactions are weak, the
nonlinear Sch\"{o}dinger equation (NLSE), also known as the Gross-Pitaevskii 
equation (GPE) provides a good description of BECs. To incorporate the statics 
and dynamical properties of TBECs, this equation can be generalized to a pair 
of CGPEs. This, however, is a description valid at zero 
temperature $T = 0$ and they form the basis of our computational scheme.
Neglecting the quantum fluctuations, the condensed state of TBEC at $T = 0$ 
can be described by the macroscopic wave function 
$\phi_1(x,y,t)$ ($\phi_2(x,y,t)$) with energy functional 
$E_{1}[\phi_1]$ ($E_{2}[\phi_2]$) for the first (second) species. The energy 
functional of the total system is
\begin{eqnarray}
E & = & E_{1} + E_{2} + E_{12}\nonumber\\
  & = & \iint dx dy\bigg[\sum_{i=1}^2\bigg(\frac{\hbar^2}{2m_i}
        |\nabla\phi_i|^2 + V_i(x,y)|\phi_i|^2
        +\frac{1}{2}U_{ii}|\phi_i|^4\bigg)\nonumber\\
        & + & U_{12}|\phi_1|^2|\phi_2|^2\bigg].
\end{eqnarray}
where $E_{12}$ is the contribution from the interspecies interaction, 
$m_i$ is the mass of the bosonic atom of species $i$, 
and $V_i (x,y)$ is the external harmonic trapping potential. The interaction 
strengths are given by $U_{ij} = 2\pi\hbar^2a_{ij}/m_{ij}$, 
where $m_{ij}^{-1} = m_{i}^{-1} + m_{j}^{-1}$ is the reduced mass 
for an atom $i$ and an atom $j$. Using these definitions and the mean-field 
theory, the static and dynamical properties of TBEC, albeit at $T = 0$, can be 
examined through the time-independent CGPE
\begin{equation}
   \left[-\frac{\hbar^2}{2m_i}\nabla^2 + V_i(x,y) + \sum_{j=1}^{2}U_{ij}|
      \phi_j|^2\right]\phi_i = \mu_i\phi_i,
\label{cgpe}
\end{equation}
which are obtained by variational minimization of the energy functional 
${\cal{E}} = E - \sum_i\mu_i N_i$ with $\phi_i^*$ as the parameter of 
variation. The Eq.~(\ref{cgpe}) forms the starting point of our analysis of 
TBECs at finite temperatures ($T\neq0$). At equilibrium, depending upon
the relative strengths of intra- ($U_{ii}$) and inter-species ($U_{12}$) 
interactions, the TBECs may either be in miscible or immiscible phase. The 
latter is also referred to as phase-separated and we use these two terms
interchangeably. The emergence of these phases renders the physics of TBEC 
drastically different from single-species BEC. And, the natural question is
the role of fluctuations, both quantum and thermal, on these phases.
For this, the first step is to solve Eqs.~(\ref{cgpe}), and then use the
HFB-Popov approximation to calculate the thermal cloud densities. 

 For $T\neq0$, along with the two coherent condensate clouds, there exist the
incoherent non-condensate clouds of both the species. This introduces additional
interparticle interactions, the intra- and inter-species interactions between 
the condensate and non-condensate clouds. The presence of larger number of 
interaction terms complicates the governing equations, and poses difficulty to 
theoretically model the system. In the present work, we have assumed that 
the thermal 
clouds of both the species are static, and consider $T$ less than the lower 
critical temperature among the two. 



\subsection{Hartree Fock Bogoliubov Theory for quasiparticle excitations }
\label{hfb_theory}

 To obtain the Hartree Fock Bogoliubov equation we consider the grand-canonical 
Hamiltonian for TBECs in a quasi-2D trap,
\begin{eqnarray}
\begin{aligned}
  \hat{H} = \sum_{i=1,2}\iint dx dy\hat{\Psi}_{i}^\dagger(x,y,t)
        \bigg[-\frac{\hbar^{2}}{2m_i}(\frac{\partial ^2}{\partial x^2} 
        + \frac{\partial ^2}{\partial y^2})
        + V_i(x,y)-\mu_i\\
        + \frac{U_{ii}}{2}\hat{\Psi}_{i}^\dagger(x,y,t)\hat{\Psi}_{i}
        (x,y,t)\bigg]\hat{\Psi}_{i}(x,y,t)\\
        + U_{12}\iint dx dy
        \hat{\Psi}_{1}^\dagger(x,y,t) \hat{\Psi}_{2}^\dagger(x,y,t)
        \hat{\Psi}_{1}(x,y,t)\hat{\Psi}_{2}(x,y,t),
\label{ham2s2d} 
\end{aligned}
\end{eqnarray}
where $i=1,2$ is the species index, $\hat{\Psi}_i$'s are the Bose field
operators of the two species, and $\mu_i$'s  are the chemical potentials.  
The intra- and interspecies interactions strengths are 
$U_{ii} = 2a_{ii}\sqrt{2\pi\lambda}$
and $U_{12}=2a_{12}\sqrt{2\pi\lambda}(1+m_1/m_2)$,
respectively, where $\lambda = (\omega_z/\omega_{\perp})$ is the
anisotropy parameter. Here, $a_{ii}$, $a_{12}$ represent the $s$-wave 
scattering 
lengths of intra and inter species interactions respectively. The requirement 
of having a quasi-2D geometry is satisfied through the following inequalities:
$\lambda \gg 1$, $\hbar\omega_z\gg\mu_i$ \cite{petrov_00, salasnich_02} and
$\hbar\omega_z\gg k_BT$ (at finite temperature $T$) 
\cite{petrov_00_1, roy_15}. Under
these constraint conditions, the motion of the trapped atoms will be confined 
strongly along $z$ direction and the atoms will remain frozen in the ground 
state providing a quasi-2D confinement.
The Heisenberg equation of motion for the Bose field 
operators $\hat{\Psi}_{i}$ in two-component notation is
\begin{equation}
 i\hbar\frac{\partial}{\partial t}
 \begin{pmatrix}
   \hat{\Psi}_1\\
   \hat{\Psi}_2
\end{pmatrix} \!\!
= \!\!\begin{pmatrix}
  \hat{h}_1 + U_{11}\hat{\Psi}_1^\dagger\hat{\Psi}_1 & U_{12}
           \hat{\Psi}_2^\dagger \hat{\Psi}_1\\
   U_{12}\hat{\Psi}_1^\dagger\hat{\Psi}_2             & \hat{h}_2
           + U_{22}\hat{\Psi}_2^\dagger \hat{\Psi}_2
  \end{pmatrix} \!\!\!
  \begin{pmatrix}
    \hat{\Psi}_1\\
    \hat{\Psi}_2  
  \end{pmatrix}, 
\label{twocomp}
\end{equation}
where $\hat{h}_{i}= (-\hbar^{2}/2m_i)(\partial ^2/\partial x^2 
+ \partial ^2/\partial y^2) + V_i(x,y)-\mu_i$. Using Bogoliubov approximation, 
the field operators can be written as
$\hat{\Psi}_i(x,y,t) = \phi_i (x,y) + \tilde\psi_i(x,y,t)$, where 
$\phi_i(x,y)$  is a $c$-field and represents the condensate, and 
$\tilde\psi_i(x,y,t) $ is the fluctuation operator corresponding to the $i$th 
species. We can 
write the total field operator as
\begin{equation}
\begin{pmatrix}
 \hat{\Psi}_1\\
 \hat{\Psi}_2
\end{pmatrix}
   = 
\begin{pmatrix}
 \phi_1\\
 \phi_2
\end{pmatrix}
 +
\begin{pmatrix}
\tilde\psi _1\\
\tilde\psi _2\\
\end{pmatrix},  
\Rightarrow \hat{\Psi} = \Phi + \tilde{\Psi},
 \label{ch2dectbec}  
\end{equation}
where $\Phi$ and $\tilde{\Psi}$ are the condensate and fluctuation
operator in two-component notations. Using the expression of $\hat{\Psi}_i$,
we can separate the Hamiltonian into terms of different orders in fluctuation
operators i.e. 
$\hat{H} = \sum_{i=1,2}\sum_{n=0}^4\hat{H}_n^i$, where $0\leqslant
n\leqslant 4$ denotes the order of the fluctuation operators. The 
explicit forms of the fluctuation operators are provided in the Appendix. 
Following the derivation,  
the equation of motion of the fluctuation operator for the first species is
\begin{eqnarray}
  i\hbar\frac{\partial \tilde{\psi_1}}{\partial t}
    &=&\left(-\frac{\hbar^2}{2m_1}\nabla^2+V_1+2U_{11}(n_{1c}
    +\tilde{n}_1)-\mu_1+ U_{12}|\phi_2|^{2}+U_{12}\tilde{n}_2\right)
    \tilde{\psi}_1\nonumber\\
    &&+U_{11}\left(\phi_{1}^{2}+\tilde{m}_1\right)\tilde{\psi}_1^\dagger
    +U_{12}\phi_1\phi_2^*\tilde{\psi}_2+U_{12}\phi_1\phi_2
    \tilde{\psi}_2^\dagger.
  \label{nceq1}
\end{eqnarray}
where for the same species $i=j$, the fluctuation operators are
$\langle\tilde{\psi}_i^\dagger\tilde{\psi}_i\rangle=\tilde{n}_i$, and
$\langle\tilde{\psi}_i\tilde{\psi}_i\rangle=\tilde{m}_i$. However, as 
mentioned 
$\langle\tilde{\psi}_i^\dagger\tilde{\psi}_j\rangle=
\langle\tilde{\psi}_i\tilde{\psi}_j\rangle=0$.

Similarly, the equation of motion of the fluctuation operator of the second
species is,
\begin{eqnarray}
  i\hbar\frac{\partial \tilde{\psi}_2}{\partial t}
    &=&\left(-\frac{\hbar^2}{2m_2}\nabla^2+V_2+2U_{22}(n_{2c}
    +\tilde{n}_2)-\mu_2 + U_{21}|\phi_1|^{2}+U_{21}\tilde{n}_1\right)
    \tilde{\psi}_2            \nonumber\\
    &&+U_{22}\left(\phi_{2}^{2}+\tilde{m}_2\right)\tilde{\psi}_2^\dagger
    +U_{21}\phi_1^*\phi_2\tilde{\psi}_1+U_{21}\phi_1\phi_2
    \tilde{\psi}_1^\dagger.
  \label{nceq2}
\end{eqnarray}
For compact notation, we have used the definitions
$n_i=n_{ic}+\tilde{n}_i$, and $m_i=\phi_{i}^{2}+\tilde{m}_i$.
The next step is to diagonalise the Hamiltonian matrix and obtain the 
quasiparticle amplitude functions $u$s and $v$s. Incorporating the Bogoliubov 
transformation, the fluctuation operators have the following form
\begin{subequations}
  \begin{eqnarray}
     \tilde{\psi}_i &=&\sum_{j}\left[u_{ij}\hat{\alpha}_j
     e^{-iE_{j}t/\hbar}-v_{ij}^{*}\hat{\alpha}_j^\dagger 
     e^{iE_{j}t/\hbar}\right],  \\
     \tilde{\psi}_{i}^\dagger &=&\sum_{j}\left[u_{ij}^*\hat{\alpha}_j^\dagger
     e^{iE_{j}t/\hbar}-v_{ij}\hat{\alpha}_j e^{-iE_{j}t/\hbar}\right].
  \end{eqnarray}
  \label{ansatzb}
\end{subequations}
Here, $j$ is the index representing the sequence of quasiparticle excitations.
We take the operators $\alpha$ and $\alpha^{\dagger}$ as common to both the 
species which is consistent in describing the coupled multispecies dynamics.
Furthermore, this reproduces the standard coupled BdG equations at $T = 0$ and 
in the limit $a_{12}\rightarrow0$, the quasiparticle spectra separates into 
two distinct sets: one set for each of the condensates. On substituting
Eq.~(\ref{ansatzb}) in Eqns.~(\ref{nceq1}) and (\ref{nceq2}) we obtain the 
BdG equations for TBEC. And, in scaled units the BdG equations are
\begin{subequations}
\begin{eqnarray}
 \hat{{\mathcal L}}_{1}u_{1j}-U_{11}\phi_{1}^{2}v_{1j}+U_{12}\phi_1 \left 
  (\phi_2^{*}u_{2j} -\phi_2v_{2j}\right )&=& E_{j}u_{1j},\;\;\;\;\;\;\\
 \hat{\underline{\mathcal L}}_{1}v_{1j}+U_{11}\phi_{1}^{*2}u_{1j}-U_{12}\phi_1^*\left (
   \phi_2v_{2j}-\phi_2^*u_{2j} \right ) &=& E_{j}v_{1j},\;\;\;\;\;\;\\
 \hat{{\mathcal L}}_{2}u_{2j}-U_{22}\phi_{2}^{2}v_{2j}+U_{12}\phi_2\left ( 
   \phi_1^*u_{1j}-\phi_1v_{1j} \right ) &=& E_{j}u_{2j},\;\;\;\;\;\;\\
 \hat{\underline{\mathcal L}}_{2}v_{2j}+U_{22}\phi_{2}^{*2}u_{2j}-U_{12} \phi_2^*\left ( 
 \phi_1v_{1j}-\phi_1^*u_{1j}\right ) &=& E_{j}v_{2j},\;\;\;\;\;\; 
\end{eqnarray}
\label{bdg}
\end{subequations}
where $\hat{{\mathcal L}}_{1}=
\big(\hat{h}_1+2U_{11}n_{1}+U_{12}n_{2})$, $\hat{{\mathcal L}}_{2}
=\big(\hat{h}_2+2U_{22}n_{2}+U_{12}n_{1}\big)$, 
$\hat{\underline{\cal L}}_i  = -\hat{\cal L}_i$, 
and the quasiparticle amplitudes are normalized as
\begin{equation}
 \iint dx dy\sum_i(|u_{ij}(x,y)|^2-|v_{ij}(x,y)|^2 = 1.
\end{equation}
Under time-independent HFB-Popov approximation for a TBEC, $\phi_i$s are 
the static solutions of the CGPEs
\begin{subequations}
\begin{eqnarray}
  \hat{h}_1\phi_1 + U_{11}\left[n_{c1}+2\tilde{n}_{1}\right]\phi_1
  +U_{12}n_2\phi_1=0,\\
  \hat{h}_2\phi_2 + U_{22}\left[n_{c2}+2\tilde{n}_{2}\right]\phi_2
  +U_{12}n_1\phi_2=0.
\end{eqnarray}
\label{gpe}
\end{subequations}
To solve Eq.~(\ref{bdg}) we define $u_{ij}$ and $v_{ij}$'s as linear
combination of $N_b$ harmonic oscillator eigenstates,
\begin{eqnarray}
\begin{aligned}
  u_{1j}(x,y)= \sum_{\kappa,l=0}^{N_b} p_{j\kappa l}\varphi_{\kappa j}(x)\varphi_{lj}(y),\;\;
  v_{1j}(x,y) = \sum_{\kappa,l=0}^{N_b} q_{j\kappa l}\varphi_{\kappa j}(x)\varphi_{lj}(y),\\
  u_{2j}(x,y) = \sum_{\kappa,l=0}^{N_b} r_{j\kappa l}\varphi_{\kappa j}(x)\varphi_{lj}(y),\;\;
  v_{2j}(x,y) = \sum_{\kappa,l=0}^{N_b} s_{j\kappa l}\varphi_{\kappa j}(x)\varphi_{lj}(y),
\label{exp}
\end{aligned}
\end{eqnarray}
where $\varphi_{kj}$s and $\varphi_{lj}$s are the $j$th harmonic oscillator 
eigenstates and
$p_{j \kappa l}$, $q_{j \kappa l}$,  $r_{j \kappa l}$ and $s_{j \kappa l}$ are 
the coefficients of linear combination. Using this expansion Eq.~(\ref{bdg}) is
reduced to a matrix eigenvalue equation and solved using standard matrix
diagonalization algorithms. The matrix has a dimension of
$4(N_b+1)\times4(N_b+1)$ and is non-Hermitian, non-symmetric 
and may have complex eigenvalues. Considering the orthogonality of harmonic 
oscillator basis, the matrix becomes sparse. Due to the $N_b^2$
scaling of the BdG matrix, the matrix size rapidly increases
with the basis size, and it is essential to use algorithms capable
of large matrix diagonalization. For this reason, we use 
\textsc{Arpack} \cite{lehoucq_98}. The eigenvalue spectrum obtained from 
the diagonalization of the matrix has an equal number of positive and negative 
eigenvalues $E_j$'s. Using the quasiparticle amplitudes obtained, the number 
density $\tilde{n}_i$ of the non-condensate atoms is
\begin{equation}
 \tilde{n}_i=\sum_{j}\{[|u_{ij}|^2+|v_{ij}|^2]N_{0}(E_j)+|v_{ij}|^2\},
 \label{n_k}
\end{equation}
where $\langle\hat{\alpha}_{j}^\dagger\hat{\alpha}_{j}\rangle = (e^{\beta
E_{j}}-1)^{-1}\equiv N_{0}(E_j)$  is the Bose factor of the quasiparticle
state with real and positive energy $E_j$. The coupled
Eqns. (\ref{bdg}) and ~(\ref{gpe}) are solved iteratively till the solutions
converge to desired accuracy. We use this theory to investigate the 
evolution of Goldstone modes and mode energies as a function of the 
interaction strengths and temperature. Although, HFB-Popov does have
the advantage vis-a-vis calculation of the modes, it is nontrivial
to get converged solutions.


\subsection{Overlap integral and dispersion relation}
A measure of phase separation is the overlap integral,
\begin{eqnarray}
  \Lambda = \frac{[\iint n_1(x,y)n_2(x,y)dxdy]^2}{[\iint n_1^2(x,y)dxdy]
          [\iint n_2^2(x,y)dxdy]}.
  \label{oint}
\end{eqnarray}
The TBEC is in the miscible phase when $\Lambda \approx 1$ and signifies 
complete overlap between the two species when $\Lambda$ has unit value. The
TBEC is completely phase separated when $\Lambda = 0$ \cite{jain_11}. 
The other important measure is the response of the TBEC when subjected to 
external perturbations, and one which defines this is the dispersion relation. 
To determine the dispersion relation we compute the root mean square of 
the wave number $k^{\rm rms}$ of each quasiparticle mode 
\cite{ticknor_14, wilson_10}
\begin{equation}
 k_j^{\rm rms} = \left\{\frac{\sum_i\int d\mathbf{k} k^2
                 [|u_{ij}(\mathbf{k})|^2 + |v_{ij}(\mathbf{k})|^2]}
                 {\sum_i\int d\mathbf{k} [|u_{ij}(\mathbf{k})|^2 
                 + |v_{ij}(\mathbf{k})|^2]}\right\}^{1/2}.
  \label{dspeq}
\end{equation}
It is to be noted here that $k_j^{\rm rms}$ are defined in terms of the
quasiparticle modes corresponding to each of the constituent species
defined in the $k$ or momentum space through the index $i = 1, 2$. It is 
then essential to compute $u_{ij}(\mathbf{k})$ and $v_{ij}(\mathbf{k})$, 
the Fourier 
transform of the Bogoliubov quasiparticle amplitudes 
$u_{ij}(x,y)$ and $v_{ij}(x,y)$,
respectively. Once we have $k_j^{\rm rms}$ for all
the modes we obtain a discrete dispersion curve. It is to be 
mentioned that to obtain $k_j^{\rm rms}$, we
consider 2D Fourier transform with ${\bf k} = (k_x, k_y)$ and the
integration in
Eq.~\ref{dspeq} is carried over in 2D Fourier space.


\subsection{Dynamical structure factor and Correlation function }
\label{dsf}

The dynamical correlation function or the dynamic structure factor (DSF) 
characterizes the dynamic properties of a quantum many body system and it is 
a quantity of considerable experimental interest. Unlike other quantum systems 
where DSF provides informations ranging from low (characterized by spectrum of 
collective excitations) to high momentum transfer (characterized by momentum 
distribution), for BECs of dilute Bose gases DSF is of importance in exploring 
the domain of high momenta, where the response of the system is not affected by 
its collective features \cite{zambelli_2000}. Rather it is determined by the 
momentum distribution of condensate atoms. In experiments DSF is measured by 
the inelastic light scattering \cite{stamper-kurn_99} and Bragg 
spectroscopy \cite{stenger_99}.  Following refs. 
\cite{zambelli_2000, wu_96, menotti_03}, the dynamic structure factor in terms 
of $j$th quasi particle amplitudes $u_{ji}(x,y)$ and $v_{ji}(x,y)$ for a TBEC is
\begin{equation}
 S_d(q_x,q_y,E)= \sum_{j,i}\Big|\iint dxdy
               [u_{ji}^{\ast}(x,y)+ v_{ji}^{\ast}(x,y)]e^{i(xq_x+yq_y)/\hbar} 
               \psi_{i}(x,y)\Big|^2\delta(E-\epsilon_j),
  \label{sfactor}
\end{equation}
where $i$ corresponds to the species index and for TBEC system $i$ = 1, 2. 
$\phi_{i}(x,y)$ is the condensate order parameter for $i$th species.

  Another important measure of the TBEC which is related to the coherence of 
the system is the first-order or the off-diagonal correlation function
\begin{eqnarray}
\label{odcond}
g_i^{(1)}(x,y,x'y') = \frac{\langle\hat{\Psi}_i^{\dagger}(x,y)
                      \hat{\Psi}_i(x',y')\rangle}
                      {\langle\hat{\Psi}_i^{\dagger}(x,y)\hat{\Psi}_i(x,y)
                      \rangle\langle\hat{\Psi}_i^{\dagger}(x',y')
                      \hat{\Psi}_i(x',y')\rangle},
\end{eqnarray}
which is also measure of the phase fluctuations. It can also be expressed in 
terms of off-diagonal condensate and noncondensate densities as
\begin{eqnarray}
\label{corr}
g_i^{(1)}(x,y,x'y') = \frac{n_{ci}(x,y;x',y')+\tilde{n}_i(x,y;x',y')}
                      {\sqrt{n_i(x,y)n_i(x',y')}},
\end{eqnarray}
where
\begin{eqnarray}
n_{ci}(x,y;x',y') &=& \phi_i^{\ast}(x,y)\phi_i(x',y'),\\
\tilde{n}_i(x,y;x',y') &=& \sum_j\{[u_{ij}^{\ast}(x,y)u_{ij}(x',y') + 
                           v_{ij}^{\ast}(x,y)v_{ij}(x',y')]N_0(E_j)
                           \nonumber\\ &+& 
                           v_{ij}^{\ast}(x,y)v_{ij}(x',y')\}
\end{eqnarray}
At $T = 0$, when the entire system is coherent and characterized
by the presence of a condensate only, then $g_i^{(1)} = 1$ within the
extent of the condensate, whether it is in the miscible or in
the immiscible regime. So, one cannot distinguish between
the two phases from the nature of the correlation functions of
the individual species. However, at $T \ne 0$, a clear signature
of a miscible-immiscible transition of the density profiles is
reflected in the form of correlation functions.


\section{Details of implementation}
\label{c_dtls}

\subsection{GPE solver and details of basis}
As a first step to compute the BdG matrix and derive the BdG equations, we 
solve the pair of CGP Eqs.~(\ref{gpe}) using split time-step 
Crank-Nicolson  \cite{muruganandam_09,vudragovic_12,young_16,young_17} and 
Fourier-pseudospectral method adapted for binary condensates. The
method when implemented with imaginary time propagation is appropriate to 
obtain the stationary ground state wave function of the TBEC. 
However, this is not the only method to solve CGP Eqs. Thorough reviews
of various numerical methods to solve GP equation, including the one we have
used, are given in refs. \cite{minguzzi_04, antoine_13}. The other numerical 
methods given in these reviews can as well be adapted to obtain the ground
state wave function of TBEC. It must also be added that a description of 
selected numerical methods to solve multicomponent BECs is reviewed in 
ref. \cite{bao_04}. To represent the quasiparticle amplitudes $u$s 
and $v$s as a linear combination of $N_b$ direct product states 
$\varphi(x)\otimes\varphi(y)$ as defined in Eq.~(\ref{exp}), $\varphi(x)$ and 
$\varphi(y)$ are considered to be the harmonic oscillator 
eigenstates~\cite{bao_05, tiwari_06}. To generate $\varphi(x)$ and 
$\varphi(y)$, we start with the ground $\varphi_0(x)$ and first excited state
$\varphi_1(x)$, and higher excited states are generated using the following 
recurrence relations
\begin{eqnarray}
 H_{n+1}(x) &=& 2xH_{n}(x) - 2nH_{n-1}(x)    \\                    
 \varphi_{n}(x) &=& \sqrt{2/n}x\varphi_{n-1}(x) - \sqrt{\frac{n-1}{n}}
                   \varphi_{n-2}(x)
\label{hpol}
\end{eqnarray} 
where $H_n(x)$ is the $n$th order Hermite polynomial. With this
choice both the CGP and BdG equations are solved using pseudospectral 
methods. The computation of basis function is implemented in the subroutine 
\texttt{basis.f90} and stored on a grid.


\subsection{BdG matrix in terms of coefficients}

The BdG matrix from the set of BdG Eqs.(\ref{bdg}) can be written as
\begin{eqnarray}
\label{bdgmat}
E
  \begin{pmatrix}
     pq \\
     rs
  \end{pmatrix}
=
 \begin{pmatrix}
  {\rm BdG}_{00} & {\rm BdG}_{10} \\
  {\rm BdG}_{01} & {\rm BdG}_{11} \\
 \end{pmatrix}
 \begin{pmatrix}
     pq \\
     rs
  \end{pmatrix},
\end{eqnarray}

where the submatrices in the above matrix equation are defined as

\begin{eqnarray}
  {\rm BdG}_{00}&=&
  \begin{pmatrix}
    \begin{matrix}
    {\mathcal{A}}_{00}&\cdots&{\mathcal{A}}_{0N_b}\\
    \vdots&\ddots&\vdots\\
    {\mathcal{A}}_{{N_b}0}&\cdots&{\mathcal{A}}_{N_b N_b}
    \end{matrix} &
    \begin{matrix}
    {\mathcal{B}}_{00}&\cdots&{\mathcal{B}}_{0N_b}\\
    \vdots&\ddots&\vdots\\
    {\mathcal{B}}_{N_b 0}&\cdots&{\mathcal{B}}_{N_b N_b}
    \end{matrix} \\
    \begin{matrix}
    {\mathcal{E}}_{00}&\cdots&{\mathcal{E}}_{0N_b}\\
    \vdots&\ddots&\vdots\\
    {\mathcal{E}}_{{N_b}0}&\cdots&{\mathcal{E}}_{N_b N_b}
    \end{matrix} &
    \begin{matrix}
    {\mathcal{F}}_{00}&\cdots&{\mathcal{F}}_{0N_b}\\
    \vdots&\ddots&\vdots\\
    {\mathcal{F}}_{N_b 0}&\cdots&{\mathcal{F}}_{N_b N_b}
    \end{matrix} 
  \end{pmatrix}, 
\end{eqnarray}
\begin{eqnarray}
  {\rm BdG}_{10}&=&
  \begin{pmatrix}
    \begin{matrix}
    {\mathcal{C}}_{00}&\cdots&{\mathcal{C}}_{0N_b}\\
    \vdots&\ddots&\vdots\\
    {\mathcal{C}}_{N_b 0}&\cdots&{\mathcal{C}}_{N_b N_b}
    \end{matrix} &
    \begin{matrix}
    {\mathcal{D}}_{00}&\cdots&{\mathcal{D}}_{0N_b}\\
    \vdots&\ddots&\vdots\\
    {\mathcal{D}}_{N_b 0}&\cdots&{\mathcal{D}}_{N_b N_b}
    \end{matrix}\\
    \begin{matrix}
    {\mathcal{G}}_{00}&\cdots&{\mathcal{G}}_{0N_b}\\
    \vdots&\ddots&\vdots\\
    {\mathcal{G}}_{N_b 0}&\cdots&{\mathcal{G}}_{N_b N_b}
    \end{matrix} &
    \begin{matrix}
    {\mathcal{H}}_{00}&\cdots&{\mathcal{H}}_{0N_b}\\
    \vdots&\ddots&\vdots\\
    {\mathcal{H}}_{N_b 0}&\cdots&{\mathcal{H}}_{N_b N_b}
    \end{matrix}
  \end{pmatrix},
\end{eqnarray}
%
\begin{eqnarray}
  {\rm BdG}_{01}&=&
  \begin{pmatrix}
    \begin{matrix}
    {\mathcal{I}}_{00}&\cdots&{\mathcal{I}}_{0N_b}\\
    \vdots&\ddots&\vdots\\
    {\mathcal{I}}_{{N_b}0}&\cdots&{\mathcal{I}}_{N_b N_b}
    \end{matrix} &
    \begin{matrix}
    {\mathcal{J}}_{00}&\cdots&{\mathcal{J}}_{0N_b}\\
    \vdots&\ddots&\vdots\\
    {\mathcal{J}}_{N_b 0}&\cdots&{\mathcal{J}}_{N_b N_b}
    \end{matrix} \\
    \begin{matrix}
    {\mathcal{M}}_{00}&\cdots&{\mathcal{M}}_{0N_b}\\
    \vdots&\ddots&\vdots\\
    {\mathcal{M}}_{{N_b}0}&\cdots&{\mathcal{M}}_{N_b N_b}
    \end{matrix} &
    \begin{matrix}
    {\mathcal{N}}_{00}&\cdots&{\mathcal{N}}_{0N_b}\\
    \vdots&\ddots&\vdots\\
    {\mathcal{N}}_{N_b 0}&\cdots&{\mathcal{N}}_{N_b N_b}
    \end{matrix} 
  \end{pmatrix}, 
\end{eqnarray}
\begin{eqnarray}
  {\rm BdG}_{11}&=&
  \begin{pmatrix}
    \begin{matrix}
    {\mathcal{K}}_{00}&\cdots&{\mathcal{K}}_{0N_b}\\
    \vdots&\ddots&\vdots\\
    {\mathcal{K}}_{N_b 0}&\cdots&{\mathcal{K}}_{N_b N_b}
    \end{matrix} &
    \begin{matrix}
    {\mathcal{L}}_{00}&\cdots&{\mathcal{L}}_{0N_b}\\
    \vdots&\ddots&\vdots\\
    {\mathcal{L}}_{N_b 0}&\cdots&{\mathcal{L}}_{N_b N_b}
    \end{matrix}\\
    \begin{matrix}
    {\mathcal{O}}_{00}&\cdots&{\mathcal{O}}_{0N_b}\\
    \vdots&\ddots&\vdots\\
    {\mathcal{O}}_{N_b 0}&\cdots&{\mathcal{O}}_{N_b N_b}
    \end{matrix} &
    \begin{matrix}
    {\mathcal{P}}_{00}&\cdots&{\mathcal{P}}_{0N_b}\\
    \vdots&\ddots&\vdots\\
    {\mathcal{P}}_{N_b 0}&\cdots&{\mathcal{P}}_{N_b N_b}
    \end{matrix}\\
  \end{pmatrix}, 
\end{eqnarray}
\begin{eqnarray}
  pq&=&
  \begin{pmatrix}
     p_{00}\\
     \vdots\\
     p_{N_bN_b}\\
     q_{00}\\
     \vdots\\
     q_{N_bN_b}
  \end{pmatrix},
\end{eqnarray}
\begin{eqnarray}
  rs&=&
  \begin{pmatrix}
     r_{00}\\
     \vdots\\
     r_{N_bN_b}\\
     s_{00}\\
     \vdots\\
     s_{N_bN_b}
  \end{pmatrix}.
\end{eqnarray}
The BdG matrix is non-Hermitian and non-symmetric with a dimension of 
$4(N_b+1)\times4(N_b+1)$, so it can have both real and complex eigenvalues 
depending on the physical parameters of the system under study.

The eigenvalue spectrum obtained from the diagonalization of the matrix has 
an equal number of positive and negative eigenvalues $E_j$'s. From the 
structure of the matrix elements, we can identify 16  blocks 
(${\mathcal{A}}$, ${\mathcal{B}}$, ${\mathcal{C}}$, ${\mathcal{D}}$, ..., 
${\mathcal{P}}$) in the BdG matrix in Eq. (\ref{bdgmat}) and in subroutine
\texttt{hfb2d2s.f90}, we compute the matrix elements for these blocks. In
subroutine \texttt{hfb2d2s.f90}, the blocks ${\mathcal{A}}$, ${\mathcal{B}}$, 
${\mathcal{C}}$, ${\mathcal{D}}$, ...,${\mathcal{P}}$ correspond to block $1,2,
3, 4,\cdots, 16$. The elements of each block have the following general 
expressions

\begin{eqnarray}
  \mathcal{A}_{p q}&=& \iint \varphi_p(x,y)[h_1 + 2U_{11}(n_{1c}+\tilde{n}_1)
                     +U_{12}(n_{2c}+\tilde{n}_2)] \varphi_q(x,y)dxdy, 
                                           \nonumber  \\
  \mathcal{B}_{p q}&=& \iint \varphi_p(x,y)[-U_{11}\phi_1^2]\varphi_q(x,y)dxdy, 
                                           \nonumber  \\
  \mathcal{C}_{p q}&=& \iint \varphi_p(x,y)[U_{12}\phi_1\phi_2^{\ast}]
                       \varphi_q(x,y)dxdy, 
                                           \nonumber  \\
  \mathcal{D}_{p q}&=& \iint \varphi_p(x,y)[-U_{12}\phi_1\phi_2]
                       \varphi_q(x,y)dxdy, 
                                           \nonumber  \\
  \mathcal{E}_{p q}&=& \iint \varphi_p(x,y)[U_{11}\phi_1^{\ast 2}]
                       \varphi_q(x,y)dxdy, 
                                           \nonumber  \\
  \mathcal{F}_{p q}&=& -\iint \varphi_p(x,y)[h_1 + 2U_{11}(n_{1c}+\tilde{n}_1)
                       +U_{12}(n_{2c}+\tilde{n}_2)] \varphi_q(x,y)dxdy,
                                           \nonumber   \\
  \mathcal{G}_{p q}&=& \iint \varphi_p(x,y)[U_{12}\phi_1^{\ast}\phi_2{\ast}]
                       \varphi_q(x,y)dxdy, 
                                           \nonumber   \\
  \mathcal{H}_{p q}&=& \iint \varphi_p(x,y)[-U_{12}\phi_1^{\ast}\phi_2]
                       \varphi_q(x,y)dxdy, 
                                           \nonumber   \\
  \mathcal{I}_{p q}&=& -\iint \varphi_p(x,y)[-U_{12}\phi_1^{\ast}\phi_2]
                       \varphi_q(x,y)dxdy,  
                                           \nonumber   \\
  \mathcal{J}_{p q}&=& \iint \varphi_p(x,y)[-U_{12}\phi_1\phi_2]
                       \varphi_q(x,y)dxdy, 
                                           \nonumber   \\
  \mathcal{K}_{p q}&=& \iint \varphi_p(x,y)[h_2 + 2U_{22}(n_{2c}+\tilde{n}_2)
                       +U_{12}(n_{1c}+\tilde{n}_1)] \varphi_q(x,y)dxdy, 
                                           \nonumber   \\
  \mathcal{L}_{p q}&=& \iint \varphi_p(x,y)[-U_{22}\phi_2^2]\varphi_q(x,y)dxdy, 
                                           \nonumber   \\
  \mathcal{M}_{p q}&=& \iint \varphi_p(x,y)[U_{12}\phi_1^{\ast}\phi_2{\ast}]
                       \varphi_q(x,y)dxdy,  
                                           \nonumber   \\
  \mathcal{N}_{p q}&=& - \iint \varphi_p(x,y)[U_{12}\phi_1\phi_2^{\ast}]
                       \varphi_q(x,y)dxdy,  
                                           \nonumber   \\
  \mathcal{O}_{p q}&=& \iint \varphi_p(x,y)[-U_{22}\phi_2^{\ast 2}]
                       \varphi_q(x,y)dxdy, 
                                           \nonumber   \\
  \mathcal{P}_{p q}&=& - \iint \varphi_p(x,y)[h_2 + 2U_{22}(n_{2c}+\tilde{n}_2)
                       +U_{12}(n_{1c}+\tilde{n}_1)] \varphi_q(x,y)dxdy 
\end{eqnarray}

The BdG matrix is  sparse as the harmonic oscillator basis are orthonormal. So,
we use sparse matrix representation to store the matrix, and diagonalized
using \textsc{Arpack} \cite{lehoucq_98} in the subroutine 
\texttt{hfbpopov.f}. Depending on the parameters, from the 
diagonalization we compute the lowest $D$ eigenvalues and corresponding $V$ 
eigenvectors.


\subsection{Computations of $u$ and $v$}
\label{uv}
From the eigenvectors of the BdG matrix, we compute the quasiparticle amplitudes
$u$ and $v$ in the subroutine \texttt{hfb2d2s.f90}. Considering the 
array of eigenvectors $V$, from Eq.~\ref{exp} the quasiparticle amplitudes are 
computed as
\begin{eqnarray}
  u_{1j}(x,y) &=& \sum_{\kappa,l=0}^{N_b}v_{n_{\kappa l}}^j
                  \varphi_{\kappa}(x) \varphi_l(y); \;\;\; 
                  0\leqslant n_{\kappa l} \leqslant (N_b+1)^2-1,\\
  v_{1j}(x,y) &=& \sum_{\kappa,l=0}^{N_b}v_{n_{\kappa l}}^j
                  \varphi_{\kappa}(x) \varphi_l(y); \;\;\;
                  (N_b+1)^2\leqslant n_{\kappa l} \leqslant 2 (N_b+1)^2-1,\\
  u_{2j}(x,y) &=& \sum_{\kappa,l=0}^{N_b}v_{n_{\kappa l}}^j
                  \varphi_{\kappa}(x)\varphi_l(y); \;\;\;
                  2(N_b+1)^2\leqslant n_{\kappa l} \leqslant 3 (N_b+1)^2-1,\\
  v_{2j}(x,y) &=& \sum_{\kappa,l=0}^{N_b}v_{n_{\kappa l}}^j
                  \varphi_{\kappa}(x)\varphi_l(y); \;\;\;
                  3(N_b+1)^2 \leqslant n_{\kappa l} \leqslant 4 (N_b+1)^2-1.
  \label{uv}
\end{eqnarray}
Here $j$ is the eigenvalue index, $v_{n_{\kappa l}}^j$ is the component of the
eigenvector and ${n_{\kappa l}}\in [0, 4 (N_b+1)^2-1]$ is the combined index to
identify the components of the eigenvectors {\em vis-a-vis} the 1D harmonic
oscillator basis.
The non-degenerate $u$s and $v$s are orthonormal. However, to make the 
degenerate $u$s and $v$s orthonormal, we use the Gram--Schmidt orthogonalization 
scheme.


\subsection{Bose factor and Goldstone modes}
Once the eigenvalues ($E_j$) of the BdG matrix are obtained after 
diagonalization, the Bose factor of the $j$th state in Eq.~(\ref{be-distbn})
is
\begin{equation}
  N_0(E_j) = \frac{1}{e^{\beta E_j}-1},
  \label{be-distbn}
\end{equation}
and the corresponding thermal or non-condensate components are computed 
using the definition of $\tilde{n}_i$ in Eq.(\ref{n_k}). As mentioned earlier,
for the degenerate states to render the $u$s and $v$s orthonormal we use the 
Gram-Schmidt orthogonalization. Among the low-energy collective modes, 
a few are 
zero energy, and these are the the Nambu-Goldstone (NG) modes. For TBEC, there 
exists two NG modes for each of the condensate species due to the breaking
of $U(1)$ global gauge symmetry when BEC is formed. These NG modes do not
contribute to $\tilde{n}_i$, and must be skipped while computing $\tilde{n}_i$.
This is implemented through the parameter \texttt{SKIP = 4} in the main 
subroutine. In the subroutine \texttt{hfb2d2s.f90},
we compute the quasi-particle amplitudes corresponding to these NG modes 
separately.

 The solutions are iterated until $n_{ic}$ and $\tilde{n}_i$ converge to a 
predefined accuracy parameter. For $T\neq0$, the convergence is either very 
slow due to the thermal fluctuations or tend to diverge. To accelerate the 
convergence and ameliorate divergence, we use the method of {\em successive
under relaxation} (SUR)\cite{simula_01}, and choose the 
underrelaxation parameter $S = 0.1$. The new solution at the 
$k$th iteration is then
\begin{equation}
  \phi_k^{\rm new}(x, y) = S\phi_k(x, y) + (1-S)\phi_{k-1}(x, y),
  \label{relax}
\end{equation}
where $k$ is the iteration index. To compute $\tilde{n}_i$ we consider the 
modes with $N_0(E_j)$ larger than a threshold value, say $10^{-3}$. For 
parameters relevant to experiments, this is achieved by considering the
first 250 or less number of modes.  

To show the structure of the code, we show a flowchart which describes the
how different modules of the code are related.
\begin{figure}[H]
\tikzstyle{decision} = [diamond, draw, fill=blue!20, 
    text width=4.5em, text badly centered, node distance=3cm, inner sep=0pt]
\tikzstyle{block} = [rectangle, draw, fill=blue!20, 
    text width=5em, text centered, rounded corners, minimum height=4em]
\tikzstyle{line} = [draw, -latex']
\tikzstyle{cloud} = [draw, ellipse,fill=red!20, node distance=3cm,
    minimum height=2em]
    
\begin{tikzpicture}[node distance = 2cm, auto]
    \node [block] (init) {Initialize $\Phi_i$};
    \node [cloud, left of=init]   (phis)  {Generate $\varphi$s};
    \node [block, below of=init]  (cggpe) {Solve coupled GGPEs};
    \node [block, below of=cggpe] (bdg)   {Evaluate BdG matrix};
    \node [cloud, right of=bdg, node distance=4cm]   (basis) 
                                          {$u$s \& $v$s in terms of 
                                           $\varphi$s};
    \node [block, below of=bdg]   (diag)  {Diagonalize BdG matrix};
    \node [cloud, right of=diag, node distance=3.3cm]   (arpack)
                                          {Use \textsc{ Arpack}};
    \node [block, left of=diag, node distance=3cm]   (iter)  
                                          {Update $\Phi_i$, $n_{ic}$ \&
                                           $\tilde{n}_i$ with SUR};
    \node [block, below of=diag]  (den)   {Compute $n_{ic}$ \& $\tilde{n}_i$};
    \node [decision, below of=den](conv) {Converged $n_{ic}$ \& $\tilde{n}_i$?};
    \node [block, below of=conv, node distance=3cm] (stop) {Stop};
    \path [line] (init) -- (cggpe);
    \path [line] (cggpe) -- (bdg);
    \path [line] (bdg) -- (diag);
    \path [line] (diag) -- (den);
    \path [line] (den) -- (conv);
    \path [line] (conv) -| node [near start] {no} (iter);
    \path [line] (iter) |- (cggpe);
    \path [line] (conv) -- node {yes}(stop);
    \path [line,dashed] (phis) -- (init);
    \path [line,dashed] (arpack) -- (diag);
    \path [line,dashed] (basis) -- (bdg);
\end{tikzpicture}
\end{figure}


\section{Description of FACt}

\subsection{Input file and parameters}

  This package requires a single input data file \texttt{input.dat}. 
It consists of ten lines, and description of the input parameters are provided 
in the contents of the sample file \texttt{input.dat} given below for 
$^{133}$Cs~-$^{87}$Rb TBEC in miscible regime shown below. 
\begin{verbatim}

280.0D0   100.0D0   100.0D0  100.0D0    !Scattering lengths  G011,G012,G021,G022
133.0D0      87.0D0                     !Masses              M1, M2
8.0D0                                   !Freq. along X drn.  NUR
1.0D0       12.5D0                      !Anisotropy          AL, LAMBDA
2000.0D0        2000.0D0                !Number of atoms     TN01, TN02
0.1D0                                   !Underrelaxation     SUNDER
55     55                               !Basis along X, Y    NBX, NBY
200    250                              !NEV NCV
0.0D-9                                  !Temperature         TEMPK
4       1                               !SKIP, ITMAX

\end{verbatim}
Where, the parameters are related to various physically significant parameters
and these are as follows:\\
\hspace*{0.5cm}
\begin{tabular} {ll}
  \texttt{G011, G022}:   & $s$-wave scattering lengths of intraspecies 
                           interaction
                           for species 1 and species 2 respectively, \\
  \texttt{G012, G021}:   & $s$-wave scattering lengths of interspecies 
                           interaction
                           between species 1 and 2, \\
  \texttt{M1, M2}    :   & Mass of species 1 and species 2 respectively,\\
  \texttt{NUR}       :   & Frequency along $x$ direction,\\  
  \texttt{AL}        :   & Anisotropy parameter in quasi-2D confinement.
                           (\texttt{AL} = $\omega_y/\omega_x$),\\
  \texttt{LAMBDA}    :   & Anisotropy parameter to create quasi-2D confinement.
                           (\texttt{LAMBDA} = $\omega_z/\omega_x$),\\
  \texttt{TN01, TN02}:   & Total number of atoms of species 1 and 2 
                           respectively, \\
  \texttt{SUNDER}    :   & Under relaxation parameter to ensure convergence, \\
  \texttt{NBX, NBY}  :   & Number of harmonic oscillator basis taken into 
                           account to construct BdG matrix,\\
  \texttt{NEV, NCV}  :   & Number of eigenvalues and eigen vectors ARPACK will
                           print in output file, \\
  \texttt{TEMPK}     :   & Temperature of the system in Kelvin,\\
  \texttt{SKIP}      :   & Number of Goldstone modes,\\ 
  \texttt{ITMAX}       :   & Number of HFB Popov self consistent iteration that
                           will ensure convergence,\\  
\end{tabular}
\newline
where, the scattering lengths are in the units of Bohr radius ($a_0$) and the 
masses are in the units of amu (atomic mass unit) 
The above sample input file corresponds to the case of radially symmetric 
\texttt{(AL = 1)} $^{133}$Cs~-$^{87}$Rb TBEC at zero temperature. To examine
the effect of anisotropy in the trapping parameters one can consider 
\texttt {AL < 1} (corresponding to $\omega_y\ll \omega_x$, the TBEC is
elongated along $y$ axis) or \texttt {AL > 1} (corresponding to 
$\omega_x\ll \omega_y$, the TBEC is elongated along $x$ axis). In our recent 
work \cite{pal_18}, we have considered the effect of anisotropy in 
$^{85}$Rb~-$^{87}$Rb TBEC at zero temperature for \texttt {AL > 1}. To make 
the system quasi-2D a large value of anisotropy parameter along axial direction
\texttt{LAMBDA = 12.5} is chosen so that the condition 
$\mu \ll\hbar\omega_z$ is satisfied. With this condition the atoms are 
strongly confined along axial ($z$) direction and they are frozen in the 
ground state. The size of the harmonic oscillator basis $\varphi_i$ chosen to 
expand $u$s and $v$s is determined by \texttt{NBX=NBY=55}. This optimal basis 
size is chosen to produce very low ($\sim O(10^{-13}$) residuals while 
diagonalising the BdG matrix using \textsc{Arpack}. Initially, the total the 
number of atoms in each species are chosen to be 2000 each 
(\texttt{TN01 = TN02 = 2000}). The under relaxation parameter 
\texttt{ SUNDER} is kept fixed at 0.1, and number of NG modes to skip 
is set to 4 (\texttt{SKIP = 4}), this avoids divergence associated with the
NG modes. The parameter \texttt{ ITMAX} is the maximum number of iterations to 
check the self consistency through HFB-Popov iterations of the BdG equations. 

  In addition to the parameters entered from the \texttt{input.dat}, there
are other parameters and variables which are defined through modules in the 
main subroutine \texttt{hfb\_main.f90}. 
The modules \texttt{COMM\_DATA}, 
\texttt{GPE\_DATA}, and \texttt{CN\_DATA} are from the original GPE solver
code \cite{muruganandam_09,vudragovic_12,young_16,young_17}. 
Solving the HFB-Popov equations 
requires additional data and variables. For this we introduce two modules 
\texttt{HFB\_2D\_DATA} and \texttt{ARPK\_DATA}. The former consists of arrays 
and constants pertaining to the BdG matrix and HFB-Popov approximation. These 
include arrays to store harmonic oscillator states $\varphi$, kinetic energy 
and potential energy contribution to BdG matrix, etc. The latter module has 
arrays and constants pertaining to \textsc{Arpack}.

\subsection{Input data}

Following input files are considered to show a testrun which takes $\approx$
10 min to complete.
 
\begin{verbatim}
280.0D0   100.0D0   100.0D0  100.0D0  !Scattering lengths  G011,G012,G021,G022
133.0D0      87.0D0                   !Masses              M1, M2
8.0D0                                 !Freq. along X drn.    NUR
1.0D0       12.5D0                    !Anisotropy          AL, LAMBDA
200.0D0        200.0D0                !Number of atoms     TN01, TN02
0.1D0                                 !Underrelaxation     SUNDER
20     20                             !Basis along X, Y    NBX, NBY
200    250                            !NEV NCV
0.0D-9                                !Temperature         TEMPK
4       1                             !SKIP, ITMAX
\end{verbatim}

\subsection{Output data}

On successful completion of computation, the package generates the eigenvalues
and eigen vectors of the BdG matrix. The eigenvalues are stored in 
data file \texttt{eigenvalue.out} and their corresponding 
quasiparticle amplitudes are stored in file \texttt{uv***.dat}. Where, 
\texttt{***} can take any value between 001 to 200. The details related to the 
computation are given in the output file \texttt{hfb2d2s.out}. Also, the eigen 
values and number of atoms at each HFB Popov iterations are written in
\texttt{hfb2d2s.out}. To check for convergence in HFB-Popov iterations, 
one needs to follow the contents of output file \texttt{converge.out}. The 
contents of the \texttt{hfb2d2s.out} file for $^{133}$Cs-~$^{87}$Rb at 
temperature $0{\rm nk}$ are written below where \texttt{Norm1} 
and \texttt{Norm2}
check the normalization, \texttt{<x1>} and \texttt{<x2>} calculate the rms sizes
or radii for species 1 and 2 respectively. \texttt{Psi1\^{}2(0)} and   
\texttt{Psi2\^{}2(0)} state the density at the center of the confining potential
for species 1 and 2 respectively.

\begin{verbatim}

------------------------------------------------------------------------------
  Trapping potential, mass and temperature of the quasi-2D TBEC
------------------------------------------------------------------------------
 ALPHA  =     1.000, LAMBDA =     12.500
 NUR    =     8.000
 M1     =   133.000, M2     =    87.000
 G011   =   280.000, G012   =   100.000
 G021   =   100.000, G022   =   100.000
 BETA   =      Infinity

------------------------------------------------------------------------------
  Derived constants, basis size and spatio-temporal grid information
------------------------------------------------------------------------------
Oscillator Length   =  0.308263D-05
MRATIO(MASS1/MASS2) =   1.529
No. of basis  X     =   20
No. of basis  Y     =   20
No of spatial points NX =   200
No of spatial points NY =   200
Spatial step size    DX = 0.050000
Spatial step size    DY = 0.050000
Temporal step size  DT = 0.001000

Total number of atoms 
TN01 =    200.00, TN02 =    200.00

Number of iterations 
NPAS =  5000 NRUN =  1000

------------------------------------------------------------------------------
 iter    Norm1    Chem1    Ener      <x1>      Psi1^2(0)   N1T
         Norm2    Chem2              <x2>      Psi2^2(0)   N2T
------------------------------------------------------------------------------
Initial :
        1.6686   1.0073   2.15965   0.94140   0.35917
        1.6686   1.1524             0.94140   0.35917

After NPAS iterations: 
        0.9941   2.9792   3.33170   1.45698   0.11648
        0.9957   2.1750             1.60807   0.10136


HFB-Popov iteration starts:
 Temp=  0.000000000000000E+000
 1      0.9941   2.9792   3.33170   1.45698   0.11648   0.00000
        0.9957   2.1750             1.60807   0.10136   0.00000
------------------------------------------------------------------------------
Eigen values correspondng to the Goldstone modes
------------------------------------------------------------------------------
nth state  real(E_n)  img(E_n)
------------------------------------------------------------------------------
  1     -0.000000    0.000000
  2     -0.000000    0.000000
  3     -0.000000    0.000000
  4     -0.000000    0.000000
------------------------------------------------------------------------------
Eigen values corresponding to quasi particle excitations
------------------------------------------------------------------------------
nth state  real(E_n)  img(E_n)
------------------------------------------------------------------------------
  6      0.057608    0.000000
  8      0.057610    0.000000
  9      0.104018    0.000000
 12      0.104018    0.000627
 13      0.134642   -0.000627
 15      0.134642    0.000000
 18      0.253197    0.000000
 19      0.696740    0.000000
 20      0.696740    0.000000
 23      0.862680    0.000000
 25      0.863303    0.000000
 27      0.868952    0.000000
 29      0.869428    0.000000
 31      0.892798    0.000000
 32      0.892798    0.000000
 35      0.893639    0.000000
 37      0.893639    0.000000
 39      0.894346    0.000000
 41      1.000080    0.000000
 44      1.000080    0.000000
 46      1.051421    0.000000
 48      1.051421    0.000000
 49      1.099959    0.000000
 51      1.099959    0.000000
 **      ********    ********

------------------------------------------------------------------------------
Scaled coupling constants and condensate atoms at each iteration
------------------------------------------------------------------------------
Iter       G11       G12         G21         G22         N01         N02
------------------------------------------------------------------------------
Initial :
       0.085195    0.038471    0.038471    0.046514      200.000000      200.000000
  1   16.991558    7.682903    7.672707    9.289328      199.443593      199.708627
 It took:   4.69246413310369      minutes.


\end{verbatim}

In the printout of the output file \texttt{hfb2d2s.out}, the rows with 
\texttt{***** } indicate the additional lines (corresponding to higher excited 
states) of data. For compactness of the manuscript, we have excluded the 
additional data of the same type. For shorter execution time of the test run 
with the above provided sample input file we have considered only one HFB-Popov
iteration. In the eigen value spectrum, the eigenvalues 
corresponding to state 12 and 13 possess imaginary part as well. These 
imaginary parts have nothing to do with the instability of the system. Rather
it is due to choice of basis size 20 which is insufficient for calculation but 
necessary for shorter execution time in testrun. \texttt{N01} and \texttt{N02}
correspond to the number of condensate atoms for species 1 and 2 respectively.
Though the eigenvalues are printed in \texttt{hfb2d2s.out}, for 
other detailed computations like the mode evolution 
as a function of anisotropy and interaction parameters, the energy eigenvalues 
are also stored in the output file \texttt{eigenvalue.out}. Such data is useful 
in studies like our previous works \cite{pal_17, pal_18}, where we have shown 
the mode evolution as a function of various parameters using this package.
It is to be mentioned that, the energy eigen 
values, chemical potentials and total energy of the system, calculated 
in this package are in units of $\hbar\omega_x$. 

\section{Numerical results}
In this section, we describe the results from our code in different
parameter regimes at zero temperature as well as in finite 
temperature. At zero temperature, the self-consistent HFB-Popov 
iterations do not produce significant changes in density profiles. Since,
HFB Popov iterations are computationally expensive and take time, the results
of zero temperature calculations are provided after single HFB-Popov iteration
(\texttt {ITMAX} = 1). Whereas for finite temperature we consider
\texttt {ITMAX} = 15 which provides required convergence.

In TBEC, the unique and easily observable effect is phase separation, where 
the density peaks of the component BECs are separate. Alternatively, we can 
say the miscible TBEC phase separates, and enters into immiscible 
configurations. Numerically, this is quantifiable from the overlap integral
$\Lambda$ as well as the quasi particle amplitudes. In two dimensional (as 
well as in quasi 2D) systems, the phase separation of TBEC can occur in two 
ways. First, the density peaks of the BECs get shifted either
along $x$-axis or along $y$-axis in $x$-$y$ plane. This type of phase 
separation is referred to as {\em side-by-side} phase separation. And second 
possibility arises when one species occupies the core region while the 
second species surrounds the first one like an annular ring.  This type  of 
phase separated density profile is termed as {\em shell structured}
density profile. In earlier kind of phase separation, the symmetry of the
confining potential is broken where as it is preserved in the latter case.

\subsection{Zero temperature} 

\begin{figure}[H]
\centering
 \includegraphics[width=10.0cm]{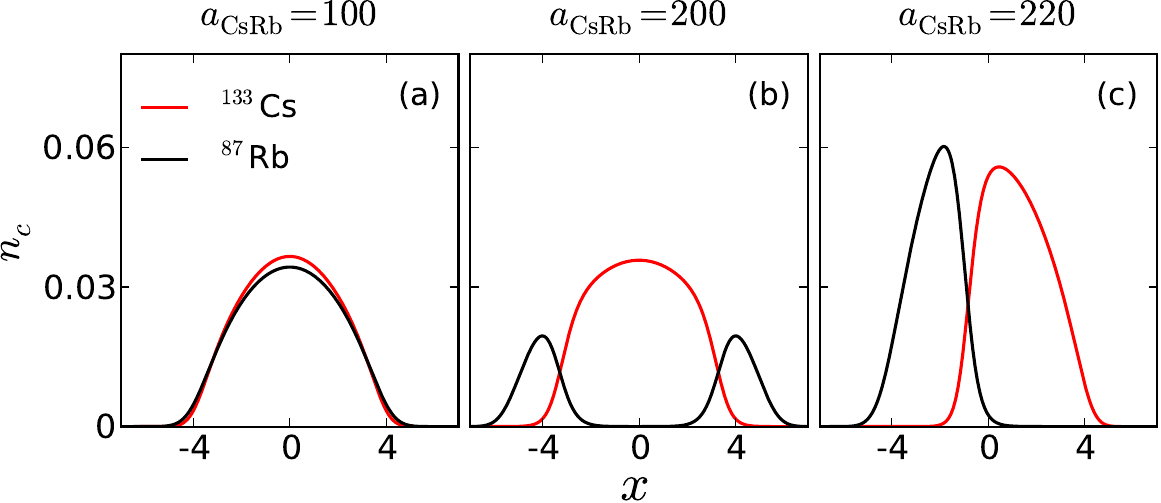}
    \caption{Equilibrium ground state of $^{133}$Cs-$^{87}$Rb TBEC at zero
 temperature for three different values of interspecies interaction strength
 (a) $a_{\rm CsRb} = 100a_0$: TBEC is in miscible domain 
(b) $a_{\rm CsRb} = 200a_0$: TBEC is in shell-structured domain  and 
(c) $a_{\rm CsRb} = 220a_0$: TBEC is side-by-side phase separated. $n_c$ is 
   measured
  in units of $a_{\rm osc}^{-2}$ and the spatial coordinate $x$ is
  measured in units of $a_{\rm osc}$.}
    \label{xcut-t0}
\end{figure}

 In this section we describe the zero temperature condensate density profiles 
$n_{ic}$ and the Bogoliubov quasi particle amplitudes $u$ and $v$ in 
miscible and immiscible regions. In Fig.~\ref{xcut-t0}, we show the density of
condensate atoms $n_{ic}(x,0)$. This figure is obtained by 
plotting column 1, 3 and 5 of file \texttt {den00x.dat} for three 
different inter species interaction strengths. If otherwise mentioned, in all 
the figures the species 1 and 2 correspond to $^{133}$Cs and $^{87}$Rb,
respectively. For Fig.\ref{xcut-t0}(a) and Fig.\ref{xcut-t0}(c) we consider 
total 2000 of atoms where as in Fig.\ref{xcut-t0}(b) we consider total 5000 
atoms. To obtain equilibrium ground states and avoid metastable states for 
side by side phase separated TBEC, it is essential to start the iterations
with the initial guess wave functions having spatially separated peaks. This 
is implemented in the subroutine \texttt{initialize.f90} by setting 
\texttt {SHIFT1} = 5.0D0. This also  ensures rapid convergence. For other 
density configurations, \texttt {SHIFT1 = 0.0D0} is considered and implies 
complete overlap of the initial guess wave functions.  
From Fig.\ref{xcut-t0}(b) it is clear that the TBEC shell-structured for the
chosen set of parameters, where $^{133}$Cs BEC is at the core and with the 
$^{87}$Rb BEC surrounding it. In Fig.\ref{xcut-t0}(c), $^{133}$Cs and
$^{87}$Rb BECs occupy right and left sides, respectively. Here, the positions 
of the BECs are not unique, and can interchange depending on the shift in
initial guess wave functions. Below we provide content of the input file to 
corresponding to Fig.\ref{xcut-t0}(a).

{\it input file corresponding to Fig.\ref{xcut-t0}(a):}
\begin{verbatim}
280.0D0   100.0D0   100.0D0  100.0D0   !Scattering lengths  G011,G012,G021,G022
133.0D0      87.0D0                    !Masses              M1, M2
8.0D0                                  !Freq. along X drn.  NUR
1.0D0       12.5D0                     !Anisotropy          AL, LAMBDA
2000.0D0        2000.0D0               !Number of atoms     TN01, TN02
0.1D0                                  !Underrelaxation     SUNDER
55     55                              !Basis along X, Y    NBX, NBY
200    250                             !NEV NCV
0.0D-9                                 !Temperature         TEMPK
4       1                              !SKIP, ITMAX
\end{verbatim}

The formation of BEC is associated with the spontaneous symmetry breaking 
(SSB) of $U(1)$ global gauge. Due to this SSB, in trapped quasi-2D TBEC, the 
low-energy BdG spectrum has two Goldstone modes for each of the condensate
species. In other words, the excitation spectrum of the BEC is gapless, and 
the two lowest energy modes with finite energies are the dipole modes. The 
dipole modes which oscillate out-of-phase with each other are called slosh 
modes. The in-phase slosh modes with center-of-mass motion are called the 
Kohn modes and have frequency identical to the natural frequency of the 
harmonic confining potential. Thus the frequency of the Kohn mode is 
independent of the type of interactions and interaction strength as well. For
this reason, getting Kohn mode energy close to 1 serves as an important
consistency check of our FACt package. 

\begin{figure}[H]
\centering
 \includegraphics[width=10.0cm]{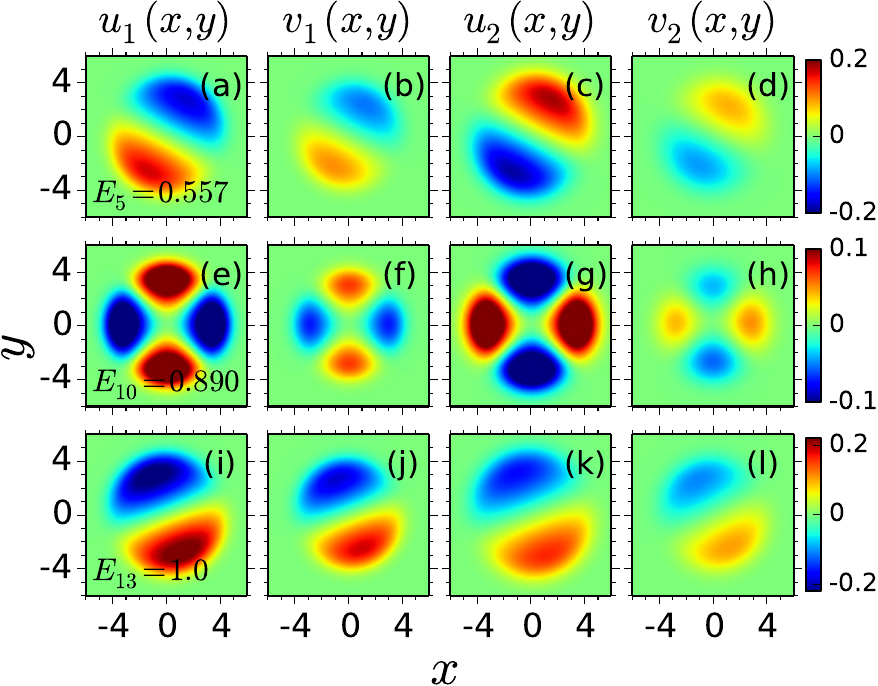}
    \caption{Quasiparticle amplitudes corresponding to miscible 
             $^{133}$Cs-$^{87}$Rb TBEC at zero temperature. (a)-(b) show slosh
             modes for species 1 and (c) - (d)
             corresponds to those of species 2.(e)-(f) show quadrupole modes
             for species 1 and (g) - (h) are those for species 2. (i)-(j) 
             describe the Kohn mode corresponding to species 1 and (k) -(l)
             are those due to species 2. $u$s and $v$s are in units of 
             $a_{\rm osc}^{-1}$ and spatial coordinate $x$ and $y$ are in 
             units of $a_{\rm osc}$.}  
    \label{qamp_mis_t0}
\end{figure}

\begin{figure}[H]
\centering
 \includegraphics[width=10.0cm]{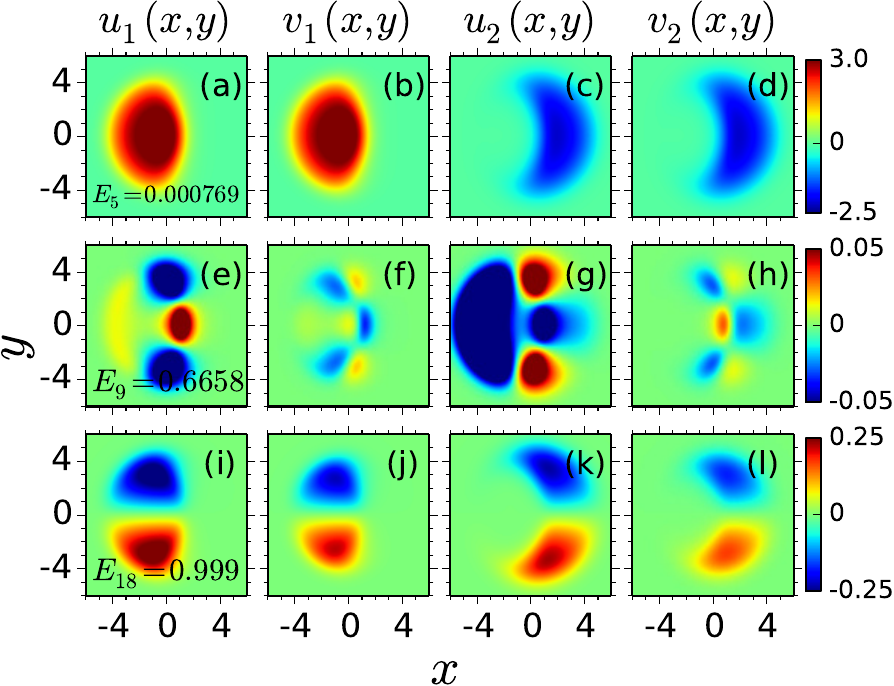}
    \caption{Quasiparticle amplitudes corresponding to side-by-side phase
             separated $^{133}$Cs-$^{87}$Rb TBEC at zero temperature. 
             (a)-(d) show quasiparticle amplitudes corresponding to NG mode
             for each of the species.(e)-(h) show those for interface mode
             for each species. (i)-(l) describe those corresponding to the Kohn
             mode for each of the species. Subscript indices 1 and 2 refer to 
             species 1 and 2 respectively. $u$s and $v$s are in units of 
             $a_{\rm osc}^{-1}$ and spatial coordinates $x$ and $y$ are in 
             units of $a_{\rm osc}$.}
    \label{qamp_sbs_t0}
\end{figure}

\begin{figure}[H]
\centering
 \includegraphics[width=10.0cm]{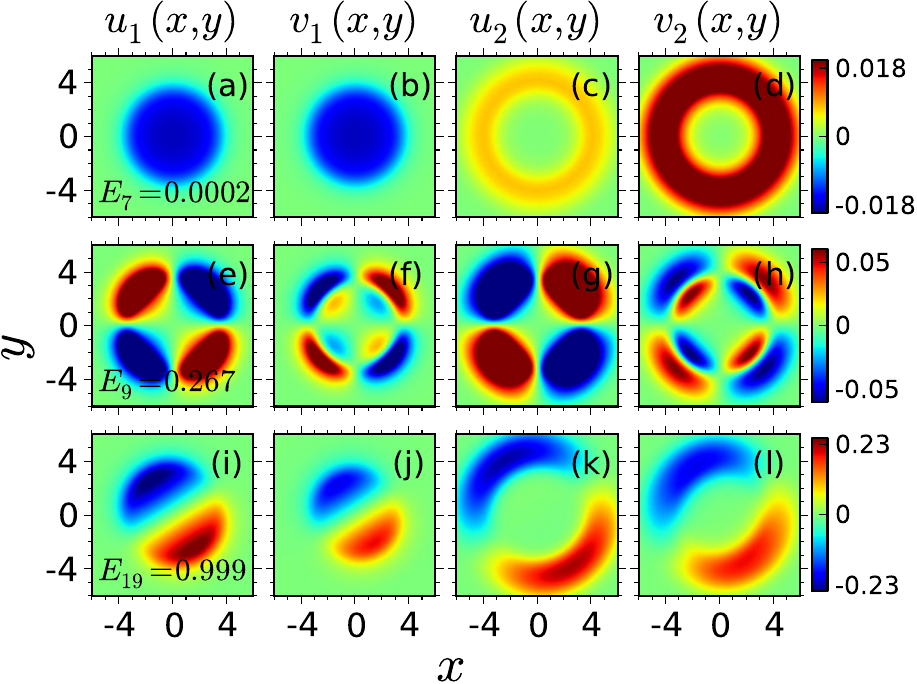}
    \caption{Quasiparticle amplitudes corresponding to shell structured
             $^{133}$Cs-$^{87}$Rb TBEC at zero temperature. (a)-(d) show 
             quasiparticle amplitudes corresponding to NG mode
             for each of the species.(e)-(h) show those for interface mode
             for each species. (i)-(l) describe those corresponding to the Kohn
             mode for
             each of the species. Like
             Fig.\ref{qamp_mis_t0} and 
             Fig.\ref{qamp_sbs_t0}, subscript indices 1 and 2 refer to 
             species 1 and 2 respectively. $u$s and $v$s are in units of 
             $a_{\rm osc}^{-1}$ and spatial coordinates $x$ and $y$ are in 
             units of $a_{\rm osc}$.}
    \label{qamp_shell_t0}
\end{figure}

The Bogoliubov quasi particle 
amplitudes corresponding to low energy modes are shown in 
Fig.~\ref{qamp_mis_t0}, \ref{qamp_sbs_t0} and \ref{qamp_shell_t0} for miscible, 
side-by-side and shell-structured TBEC respectively.
 
The quasiparticle amplitudes of the selected low-energy modes in the 
miscible domain obtained with $a_{\rm CsRb} = 100a_0$ are shown in 
Fig.~\ref{qamp_mis_t0}. The images in Fig.~\ref{qamp_mis_t0} (a)-(d) correspond 
to the slosh mode of the system. To obtain the quasiparticle amplitudes, we 
plot column 3, 4, 5 and 6 of file \texttt{uv005.dat}. In 
Fig.~\ref{qamp_mis_t0}(e)-(h), the quasiparticle amplitudes from the file 
\texttt{uv010.dat} are shown, and these correspond to quadrupole mode of the 
system. And, the Kohn modes, from the data in the file \texttt{uv013.dat}, are 
shown in Fig.~\ref{qamp_mis_t0}(i)-(l). Here, the numerical 
value \texttt{013} in
file name \texttt{uv013.dat} indicates that it is the 13th excited state. For 
each of the quasiparticle amplitudes the corresponding energies, taken from 
the output file \texttt{eigenvalue.dat}, are given in the bottom left corner. 
 
For the case of side-by-side immiscible phase, with $a_{\rm CsRb} = 220a_0$,
the quasiparticle amplitudes of low-lying modes are shown in 
Fig.~\ref{qamp_sbs_t0}. The images in Fig.~\ref{qamp_sbs_t0} (a)-(d) correspond 
to the NG modes of the system which in general resemble $n_{ic}$, and are based
on the data in the output file \texttt{uv005.dat}. Due to the rotational 
symmetry breaking associated with the miscible to side-by-side immiscible 
phase transition, each species has two additional NG modes. The 
Fig.\ref{qamp_sbs_t0}(e)-(h) show the quasiparticle amplitudes from 
\texttt{uv009.dat}, and these correspond to interface mode of the
system. In the immiscible domain the interface modes, as the name suggests,
are localized at the interface of the two species. The Kohn modes of the 
system are shown in Fig.~\ref{qamp_sbs_t0}(i)-(l) which correspond to the data
in \texttt{uv018.dat}. 
  
For shell-structured TBEC, the quasiparticle amplitudes corresponding to 
NG modes, quadrupole modes and Kohn modes are shown in 
Fig.\ref{qamp_shell_t0}(a)-(d), (e)-(h) and (i)-(l) respectively.

\subsection{Finite temperature}

\begin{figure}[H]
\centering
 \includegraphics[width=10.0cm]{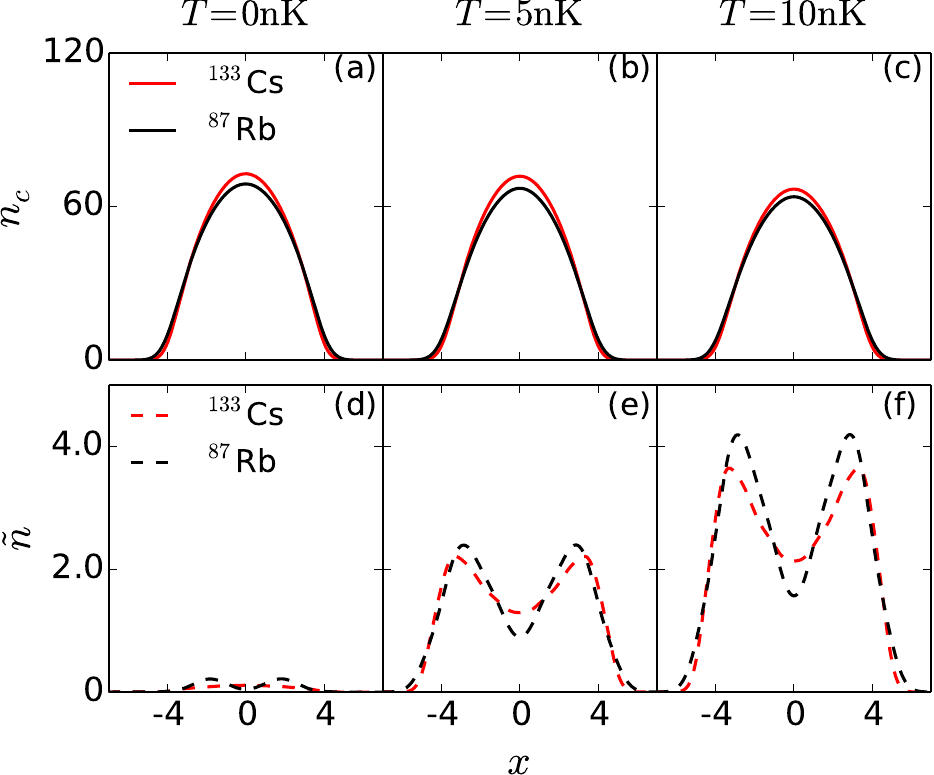}
    \caption{Equilibrium ground state density of $^{133}$Cs-$^{87}$Rb TBEC 
             in miscible domain for three different values of temperature
             (a) $T = 0 {\rm nK}$ (b) $T = 5 {\rm nK}$ 
             and (c) $T = 10{\rm nK}$. Interspecies 
             interaction strength is fixed at $a_{\rm CsRb} = 100a_0$. $n_c$ and
             $\tilde{n}$ are measured in units of $a_{\rm osc}^{-2}$ and the 
             spatial coordinate $x$ is measured in units of $a_{\rm osc}$. }
    \label{den-finite-t}
\end{figure}

\begin{figure}[H]
\centering
 \includegraphics[width=10.0cm]{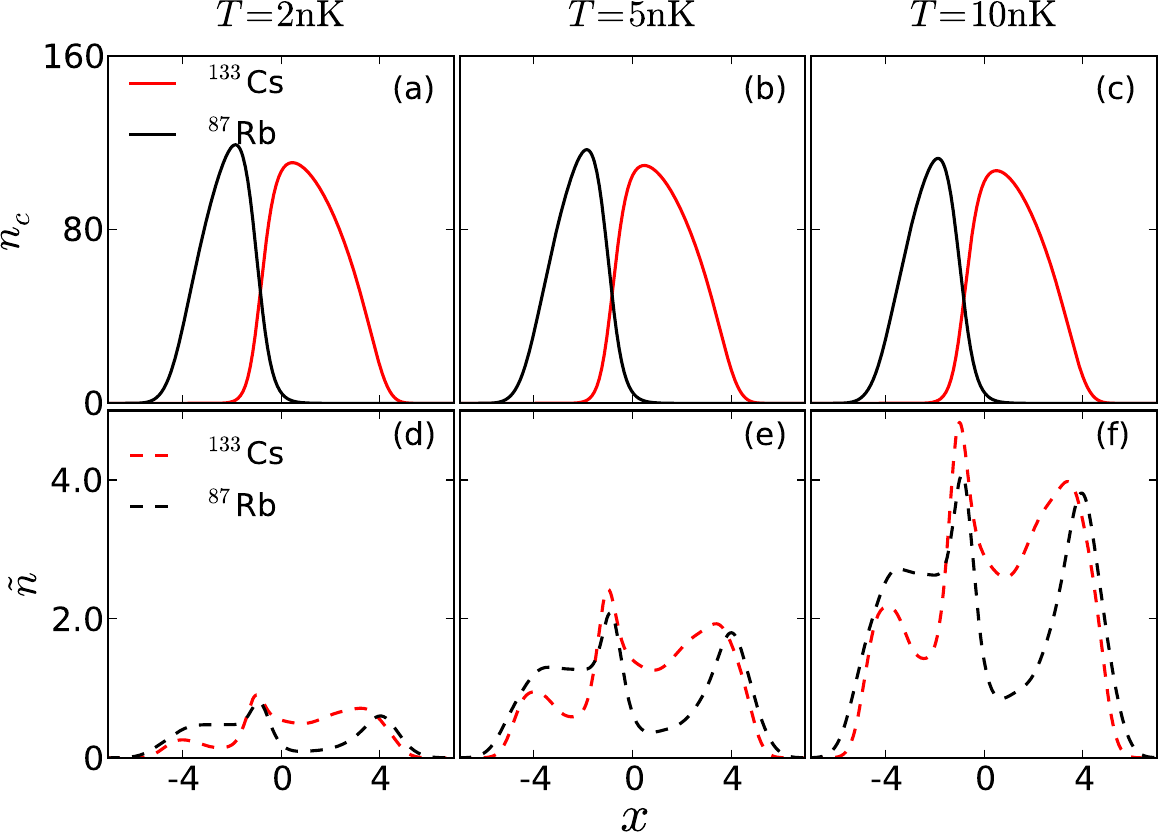}
    \caption{Equilibrium ground state density of $^{133}$Cs-$^{87}$Rb TBEC 
             in immiscible (side-by-side) domain for three different values
             of temperature (a) $T = 2{\rm nK}$ (b) $T = 5{\rm nK}$ 
             and (c) $T = 10{\rm nK}$. 
             Interspecies interaction strength is fixed at $a_{\rm CsRb} 
             = 220a_0$.
             $n_c$ and $\tilde{n}$ are measured in units of 
             $a_{\rm osc}^{-2}$ and the spatial coordinate $x$ is 
             measured in units of $a_{\rm osc}$.}
    \label{sbs-finite-t}
\end{figure}

For finite temperature computations, solving the HFB-Popov equations require 
iterations and we consider \texttt{ ITMAX} = 15 for all the finite temperature 
computations reported in this work. The density profiles of $n_{ic}$ 
corresponding to each HFB-Popov iterations are stored in the file 
\texttt{den00x.dat} where \texttt{ x} runs from 0 to ITMAX. When $T\neq 0$, at 
each iteration, the number of condensate atoms decreases, whereas the number 
of thermal (non condensate) atoms increases. Fig.~\ref{den-finite-t} shows the 
equilibrium profiles of $n_{ic}$ and $\tilde{n}_{ic}$ for three different 
temperatures in miscible domain. The plots in Fig.~\ref{den-finite-t}(a) 
correspond to $n_{ic}$ at $T = 0 {\rm nK}$, and hence in 
Fig.~\ref{den-finite-t}(d)
$\tilde{n}_{ic}$ are negligibly small. The plots in 
Fig.~\ref{den-finite-t}(b) and (c) correspond to $n_{ic}$ at $T = 5 {\rm nK}$ 
and 
$T = 10 {\rm nK}$, respectively. To obtain the plots in the top row, we plotted 
column 1, column 3 and column 5 file of \texttt{den00x.dat} with column 3 
and column 5 multiplied by number of condensate atoms $N_{01}$ and $N_{02}$
(taken from \texttt{hfb2d2s.out}), respectively. Although, the changes in 
$n_{ic}$ are not dramatic,  there is a large change in  $\tilde{n}_{ic}$
as shown in Fig.\ref{den-finite-t}(e)-(f). From Fig.~\ref{den-finite-t},
there is a notable feature of $\tilde{n}_{ic}$: it has a minimum where
$n_{ic}$ has maximum value.
For the side-by-side configuration the density profiles at finite temperature
are shown in Fig.~\ref{sbs-finite-t}. Like in the miscible domain, here 
as well, we observe growth in $\tilde{n}_{ic}$ with the increase of temperature 
and thereby lowering the number of condensate atoms. It is to be noted that at 
the interface of two species, where the $n_{ic}$ are low, $\tilde{n}_{ic}$
have maximum value.  

\section*{Acknowledgments}
The example results shown
in the paper are based on the computations using the HPC cluster Vikram-100
at Physical Research Laboratory, Ahmedabad and Max Planck Computing Facility. 
S. G. acknowledges the support of the 
Science \& Engineering Research Board (SERB), Department of Science and Technology, 
Government of India under the project ECR/2017/001436 and 
ISIRD project 9-256/2016/IITRPR/823 of Indian
Institute of Technology (IIT) Ropar.

\bibliographystyle{apsrev4-1}
\bibliography{references}

\begin{thebibliography}{50}%
\makeatletter
\providecommand \@ifxundefined [1]{%
 \@ifx{#1\undefined}
}%
\providecommand \@ifnum [1]{%
 \ifnum #1\expandafter \@firstoftwo
 \else \expandafter \@secondoftwo
 \fi
}%
\providecommand \@ifx [1]{%
 \ifx #1\expandafter \@firstoftwo
 \else \expandafter \@secondoftwo
 \fi
}%
\providecommand \natexlab [1]{#1}%
\providecommand \enquote  [1]{``#1''}%
\providecommand \bibnamefont  [1]{#1}%
\providecommand \bibfnamefont [1]{#1}%
\providecommand \citenamefont [1]{#1}%
\providecommand \href@noop [0]{\@secondoftwo}%
\providecommand \href [0]{\begingroup \@sanitize@url \@href}%
\providecommand \@href[1]{\@@startlink{#1}\@@href}%
\providecommand \@@href[1]{\endgroup#1\@@endlink}%
\providecommand \@sanitize@url [0]{\catcode `\\12\catcode `\$12\catcode
  `\&12\catcode `\#12\catcode `\^12\catcode `\_12\catcode `\%12\relax}%
\providecommand \@@startlink[1]{}%
\providecommand \@@endlink[0]{}%
\providecommand \url  [0]{\begingroup\@sanitize@url \@url }%
\providecommand \@url [1]{\endgroup\@href {#1}{\urlprefix }}%
\providecommand \urlprefix  [0]{URL }%
\providecommand \Eprint [0]{\href }%
\providecommand \doibase [0]{http://dx.doi.org/}%
\providecommand \selectlanguage [0]{\@gobble}%
\providecommand \bibinfo  [0]{\@secondoftwo}%
\providecommand \bibfield  [0]{\@secondoftwo}%
\providecommand \translation [1]{[#1]}%
\providecommand \BibitemOpen [0]{}%
\providecommand \bibitemStop [0]{}%
\providecommand \bibitemNoStop [0]{.\EOS\space}%
\providecommand \EOS [0]{\spacefactor3000\relax}%
\providecommand \BibitemShut  [1]{\csname bibitem#1\endcsname}%
\let\auto@bib@innerbib\@empty
\bibitem [{\citenamefont {Griffin}(1996)}]{griffin_96}%
  \BibitemOpen
  \bibfield  {author} {\bibinfo {author} {\bibfnamefont {A.}~\bibnamefont
  {Griffin}},\ }\href {\doibase 10.1103/PhysRevB.53.9341} {\bibfield  {journal}
  {\bibinfo  {journal} {Phys. Rev. B}\ }\textbf {\bibinfo {volume} {53}},\
  \bibinfo {pages} {9341} (\bibinfo {year} {1996})}\BibitemShut {NoStop}%
\bibitem [{\citenamefont {Hutchinson}\ \emph {et~al.}(1997)\citenamefont
  {Hutchinson}, \citenamefont {Zaremba},\ and\ \citenamefont
  {Griffin}}]{hutchinson_97}%
  \BibitemOpen
  \bibfield  {author} {\bibinfo {author} {\bibfnamefont {D.~A.~W.}\
  \bibnamefont {Hutchinson}}, \bibinfo {author} {\bibfnamefont
  {E.}~\bibnamefont {Zaremba}}, \ and\ \bibinfo {author} {\bibfnamefont
  {A.}~\bibnamefont {Griffin}},\ }\href {\doibase 10.1103/PhysRevLett.78.1842}
  {\bibfield  {journal} {\bibinfo  {journal} {Phys. Rev. Lett.}\ }\textbf
  {\bibinfo {volume} {78}},\ \bibinfo {pages} {1842} (\bibinfo {year}
  {1997})}\BibitemShut {NoStop}%
\bibitem [{\citenamefont {Dodd}\ \emph {et~al.}(1998)\citenamefont {Dodd},
  \citenamefont {Edwards}, \citenamefont {Clark},\ and\ \citenamefont
  {Burnett}}]{dodd_98}%
  \BibitemOpen
  \bibfield  {author} {\bibinfo {author} {\bibfnamefont {R.~J.}\ \bibnamefont
  {Dodd}}, \bibinfo {author} {\bibfnamefont {M.}~\bibnamefont {Edwards}},
  \bibinfo {author} {\bibfnamefont {C.~W.}\ \bibnamefont {Clark}}, \ and\
  \bibinfo {author} {\bibfnamefont {K.}~\bibnamefont {Burnett}},\ }\href
  {\doibase 10.1103/PhysRevA.57.R32} {\bibfield  {journal} {\bibinfo  {journal}
  {Phys. Rev. A}\ }\textbf {\bibinfo {volume} {57}},\ \bibinfo {pages} {R32}
  (\bibinfo {year} {1998})}\BibitemShut {NoStop}%
\bibitem [{\citenamefont {Gies}\ \emph {et~al.}(2004)\citenamefont {Gies},
  \citenamefont {van Zyl}, \citenamefont {Morgan},\ and\ \citenamefont
  {Hutchinson}}]{gies_04}%
  \BibitemOpen
  \bibfield  {author} {\bibinfo {author} {\bibfnamefont {C.}~\bibnamefont
  {Gies}}, \bibinfo {author} {\bibfnamefont {B.~P.}\ \bibnamefont {van Zyl}},
  \bibinfo {author} {\bibfnamefont {S.~A.}\ \bibnamefont {Morgan}}, \ and\
  \bibinfo {author} {\bibfnamefont {D.~A.~W.}\ \bibnamefont {Hutchinson}},\
  }\href {\doibase 10.1103/PhysRevA.69.023616} {\bibfield  {journal} {\bibinfo
  {journal} {Phys. Rev. A}\ }\textbf {\bibinfo {volume} {69}},\ \bibinfo
  {pages} {023616} (\bibinfo {year} {2004})}\BibitemShut {NoStop}%
\bibitem [{\citenamefont {Jin}\ \emph {et~al.}(1997)\citenamefont {Jin},
  \citenamefont {Matthews}, \citenamefont {Ensher}, \citenamefont {Wieman},\
  and\ \citenamefont {Cornell}}]{jin_97}%
  \BibitemOpen
  \bibfield  {author} {\bibinfo {author} {\bibfnamefont {D.~S.}\ \bibnamefont
  {Jin}}, \bibinfo {author} {\bibfnamefont {M.~R.}\ \bibnamefont {Matthews}},
  \bibinfo {author} {\bibfnamefont {J.~R.}\ \bibnamefont {Ensher}}, \bibinfo
  {author} {\bibfnamefont {C.~E.}\ \bibnamefont {Wieman}}, \ and\ \bibinfo
  {author} {\bibfnamefont {E.~A.}\ \bibnamefont {Cornell}},\ }\href {\doibase
  10.1103/PhysRevLett.78.764} {\bibfield  {journal} {\bibinfo  {journal} {Phys.
  Rev. Lett.}\ }\textbf {\bibinfo {volume} {78}},\ \bibinfo {pages} {764}
  (\bibinfo {year} {1997})}\BibitemShut {NoStop}%
\bibitem [{\citenamefont {Gautam}\ and\ \citenamefont
  {Angom}(2011)}]{gautam_11}%
  \BibitemOpen
  \bibfield  {author} {\bibinfo {author} {\bibfnamefont {S.}~\bibnamefont
  {Gautam}}\ and\ \bibinfo {author} {\bibfnamefont {D.}~\bibnamefont {Angom}},\
  }\href {\doibase doi:10.1088/0953-4075/44/2/025302} {\bibfield  {journal}
  {\bibinfo  {journal} {Journal of Physics B: Atomic, Molecular and Optical
  Physics}\ }\textbf {\bibinfo {volume} {44}},\ \bibinfo {pages} {025302}
  (\bibinfo {year} {2011})}\BibitemShut {NoStop}%
\bibitem [{\citenamefont {Gautam}\ and\ \citenamefont
  {Angom}(2010)}]{gautam_10}%
  \BibitemOpen
  \bibfield  {author} {\bibinfo {author} {\bibfnamefont {S.}~\bibnamefont
  {Gautam}}\ and\ \bibinfo {author} {\bibfnamefont {D.}~\bibnamefont {Angom}},\
  }\href {\doibase 10.1103/PhysRevA.81.053616} {\bibfield  {journal} {\bibinfo
  {journal} {Phys. Rev. A}\ }\textbf {\bibinfo {volume} {81}},\ \bibinfo
  {pages} {053616} (\bibinfo {year} {2010})}\BibitemShut {NoStop}%
\bibitem [{\citenamefont {Pires}\ and\ \citenamefont
  {de~Passos}(2008)}]{pires_08}%
  \BibitemOpen
  \bibfield  {author} {\bibinfo {author} {\bibfnamefont {M.~O.~C.}\
  \bibnamefont {Pires}}\ and\ \bibinfo {author} {\bibfnamefont {E.~J.~V.}\
  \bibnamefont {de~Passos}},\ }\href {\doibase 10.1103/PhysRevA.77.033606}
  {\bibfield  {journal} {\bibinfo  {journal} {Phys. Rev. A}\ }\textbf {\bibinfo
  {volume} {77}},\ \bibinfo {pages} {033606} (\bibinfo {year}
  {2008})}\BibitemShut {NoStop}%
\bibitem [{\citenamefont {Roy}\ \emph {et~al.}(2014)\citenamefont {Roy},
  \citenamefont {Gautam},\ and\ \citenamefont {Angom}}]{roy_14}%
  \BibitemOpen
  \bibfield  {author} {\bibinfo {author} {\bibfnamefont {A.}~\bibnamefont
  {Roy}}, \bibinfo {author} {\bibfnamefont {S.}~\bibnamefont {Gautam}}, \ and\
  \bibinfo {author} {\bibfnamefont {D.}~\bibnamefont {Angom}},\ }\href
  {\doibase 10.1103/PhysRevA.89.013617} {\bibfield  {journal} {\bibinfo
  {journal} {Phys. Rev. A}\ }\textbf {\bibinfo {volume} {89}},\ \bibinfo
  {pages} {013617} (\bibinfo {year} {2014})}\BibitemShut {NoStop}%
\bibitem [{\citenamefont {Roy}\ and\ \citenamefont {Angom}(2015)}]{roy_15}%
  \BibitemOpen
  \bibfield  {author} {\bibinfo {author} {\bibfnamefont {A.}~\bibnamefont
  {Roy}}\ and\ \bibinfo {author} {\bibfnamefont {D.}~\bibnamefont {Angom}},\
  }\href {\doibase 10.1103/PhysRevA.92.011601} {\bibfield  {journal} {\bibinfo
  {journal} {Phys. Rev. A}\ }\textbf {\bibinfo {volume} {92}},\ \bibinfo
  {pages} {011601} (\bibinfo {year} {2015})}\BibitemShut {NoStop}%
\bibitem [{\citenamefont {Roy}\ and\ \citenamefont {Angom}(2016)}]{roy_16}%
  \BibitemOpen
  \bibfield  {author} {\bibinfo {author} {\bibfnamefont {A.}~\bibnamefont
  {Roy}}\ and\ \bibinfo {author} {\bibfnamefont {D.}~\bibnamefont {Angom}},\
  }\href {\doibase 10.1088/1367-2630/18/8/083007} {\bibfield  {journal}
  {\bibinfo  {journal} {New Journal of Physics}\ }\textbf {\bibinfo {volume}
  {18}},\ \bibinfo {pages} {083007} (\bibinfo {year} {2016})}\BibitemShut
  {NoStop}%
\bibitem [{\citenamefont {Cabrera}\ \emph {et~al.}(2018)\citenamefont
  {Cabrera}, \citenamefont {Tanzi}, \citenamefont {Sanz}, \citenamefont
  {Naylor}, \citenamefont {Thomas}, \citenamefont {Cheiney},\ and\
  \citenamefont {Tarruell}}]{cabrera_18}%
  \BibitemOpen
  \bibfield  {author} {\bibinfo {author} {\bibfnamefont {C.~R.}\ \bibnamefont
  {Cabrera}}, \bibinfo {author} {\bibfnamefont {L.}~\bibnamefont {Tanzi}},
  \bibinfo {author} {\bibfnamefont {J.}~\bibnamefont {Sanz}}, \bibinfo {author}
  {\bibfnamefont {B.}~\bibnamefont {Naylor}}, \bibinfo {author} {\bibfnamefont
  {P.}~\bibnamefont {Thomas}}, \bibinfo {author} {\bibfnamefont
  {P.}~\bibnamefont {Cheiney}}, \ and\ \bibinfo {author} {\bibfnamefont
  {L.}~\bibnamefont {Tarruell}},\ }\href {\doibase 10.1126/science.aao5686}
  {\bibfield  {journal} {\bibinfo  {journal} {Science}\ }\textbf {\bibinfo
  {volume} {359}},\ \bibinfo {pages} {301} (\bibinfo {year}
  {2018})}\BibitemShut {NoStop}%
\bibitem [{\citenamefont {Petrov}(2015)}]{petrov_15}%
  \BibitemOpen
  \bibfield  {author} {\bibinfo {author} {\bibfnamefont {D.~S.}\ \bibnamefont
  {Petrov}},\ }\href {\doibase 10.1103/PhysRevLett.115.155302} {\bibfield
  {journal} {\bibinfo  {journal} {Phys. Rev. Lett.}\ }\textbf {\bibinfo
  {volume} {115}},\ \bibinfo {pages} {155302} (\bibinfo {year}
  {2015})}\BibitemShut {NoStop}%
\bibitem [{\citenamefont {Petrov}\ and\ \citenamefont
  {Astrakharchik}(2016)}]{petrov_16}%
  \BibitemOpen
  \bibfield  {author} {\bibinfo {author} {\bibfnamefont {D.~S.}\ \bibnamefont
  {Petrov}}\ and\ \bibinfo {author} {\bibfnamefont {G.~E.}\ \bibnamefont
  {Astrakharchik}},\ }\href {\doibase 10.1103/PhysRevLett.117.100401}
  {\bibfield  {journal} {\bibinfo  {journal} {Phys. Rev. Lett.}\ }\textbf
  {\bibinfo {volume} {117}},\ \bibinfo {pages} {100401} (\bibinfo {year}
  {2016})}\BibitemShut {NoStop}%
\bibitem [{\citenamefont {Semeghini}\ \emph {et~al.}(2018)\citenamefont
  {Semeghini}, \citenamefont {Ferioli}, \citenamefont {Masi}, \citenamefont
  {Mazzinghi}, \citenamefont {Wolswijk}, \citenamefont {Minardi}, \citenamefont
  {Modugno}, \citenamefont {Modugno}, \citenamefont {Inguscio},\ and\
  \citenamefont {Fattori}}]{semeghini_18}%
  \BibitemOpen
  \bibfield  {author} {\bibinfo {author} {\bibfnamefont {G.}~\bibnamefont
  {Semeghini}}, \bibinfo {author} {\bibfnamefont {G.}~\bibnamefont {Ferioli}},
  \bibinfo {author} {\bibfnamefont {L.}~\bibnamefont {Masi}}, \bibinfo {author}
  {\bibfnamefont {C.}~\bibnamefont {Mazzinghi}}, \bibinfo {author}
  {\bibfnamefont {L.}~\bibnamefont {Wolswijk}}, \bibinfo {author}
  {\bibfnamefont {F.}~\bibnamefont {Minardi}}, \bibinfo {author} {\bibfnamefont
  {M.}~\bibnamefont {Modugno}}, \bibinfo {author} {\bibfnamefont
  {G.}~\bibnamefont {Modugno}}, \bibinfo {author} {\bibfnamefont
  {M.}~\bibnamefont {Inguscio}}, \ and\ \bibinfo {author} {\bibfnamefont
  {M.}~\bibnamefont {Fattori}},\ }\href {\doibase
  10.1103/PhysRevLett.120.235301} {\bibfield  {journal} {\bibinfo  {journal}
  {Phys. Rev. Lett.}\ }\textbf {\bibinfo {volume} {120}},\ \bibinfo {pages}
  {235301} (\bibinfo {year} {2018})}\BibitemShut {NoStop}%
\bibitem [{\citenamefont {Pal}\ \emph {et~al.}(2017)\citenamefont {Pal},
  \citenamefont {Roy},\ and\ \citenamefont {Angom}}]{pal_17}%
  \BibitemOpen
  \bibfield  {author} {\bibinfo {author} {\bibfnamefont {S.}~\bibnamefont
  {Pal}}, \bibinfo {author} {\bibfnamefont {A.}~\bibnamefont {Roy}}, \ and\
  \bibinfo {author} {\bibfnamefont {D.}~\bibnamefont {Angom}},\ }\href
  {\doibase 10.1088/1361-6455/aa889f} {\bibfield  {journal} {\bibinfo
  {journal} {J. Phys B.}\ }\textbf {\bibinfo {volume} {50}},\ \bibinfo {pages}
  {195301} (\bibinfo {year} {2017})}\BibitemShut {NoStop}%
\bibitem [{\citenamefont {Pal}\ \emph {et~al.}(2018)\citenamefont {Pal},
  \citenamefont {Roy},\ and\ \citenamefont {Angom}}]{pal_18}%
  \BibitemOpen
  \bibfield  {author} {\bibinfo {author} {\bibfnamefont {S.}~\bibnamefont
  {Pal}}, \bibinfo {author} {\bibfnamefont {A.}~\bibnamefont {Roy}}, \ and\
  \bibinfo {author} {\bibfnamefont {D.}~\bibnamefont {Angom}},\ }\href
  {\doibase 10.1088/1361-6455/aab541} {\bibfield  {journal} {\bibinfo
  {journal} {J. Phys. B}\ }\textbf {\bibinfo {volume} {51}},\ \bibinfo {pages}
  {085302} (\bibinfo {year} {2018})}\BibitemShut {NoStop}%
\bibitem [{\citenamefont {Hohenberg}\ and\ \citenamefont
  {Martin}(1965)}]{hohenberg_65}%
  \BibitemOpen
  \bibfield  {author} {\bibinfo {author} {\bibfnamefont {P.}~\bibnamefont
  {Hohenberg}}\ and\ \bibinfo {author} {\bibfnamefont {P.}~\bibnamefont
  {Martin}},\ }\href {\doibase 10.1016/0003-4916(65)90280-0} {\bibfield
  {journal} {\bibinfo  {journal} {Annals of Physics}\ }\textbf {\bibinfo
  {volume} {34}},\ \bibinfo {pages} {291} (\bibinfo {year} {1965})}\BibitemShut
  {NoStop}%
\bibitem [{\citenamefont {Mermin}\ and\ \citenamefont
  {Wagner}(1966)}]{mermin_66}%
  \BibitemOpen
  \bibfield  {author} {\bibinfo {author} {\bibfnamefont {N.~D.}\ \bibnamefont
  {Mermin}}\ and\ \bibinfo {author} {\bibfnamefont {H.}~\bibnamefont
  {Wagner}},\ }\href {\doibase 10.1103/PhysRevLett.17.1133} {\bibfield
  {journal} {\bibinfo  {journal} {Phys. Rev. Lett.}\ }\textbf {\bibinfo
  {volume} {17}},\ \bibinfo {pages} {1133} (\bibinfo {year}
  {1966})}\BibitemShut {NoStop}%
\bibitem [{\citenamefont {Berezinskii}(1972)}]{berezinskii_72}%
  \BibitemOpen
  \bibfield  {author} {\bibinfo {author} {\bibfnamefont {V.~.~L.}\ \bibnamefont
  {Berezinskii}},\ }\href@noop {} {\bibfield  {journal} {\bibinfo  {journal}
  {Sov. Phys. JETP}\ }\textbf {\bibinfo {volume} {34}},\ \bibinfo {pages} {610}
  (\bibinfo {year} {1972})}\BibitemShut {NoStop}%
\bibitem [{\citenamefont {Kosterlitz}\ and\ \citenamefont
  {Thouless}(1972)}]{kosterlitz_72}%
  \BibitemOpen
  \bibfield  {author} {\bibinfo {author} {\bibfnamefont {J.~M.}\ \bibnamefont
  {Kosterlitz}}\ and\ \bibinfo {author} {\bibfnamefont {D.~J.}\ \bibnamefont
  {Thouless}},\ }\href {http://stacks.iop.org/0022-3719/5/i=11/a=002}
  {\bibfield  {journal} {\bibinfo  {journal} {Journal of Physics C: Solid State
  Physics}\ }\textbf {\bibinfo {volume} {5}},\ \bibinfo {pages} {L124}
  (\bibinfo {year} {1972})}\BibitemShut {NoStop}%
\bibitem [{\citenamefont {Kosterlitz}\ and\ \citenamefont
  {Thouless}(1973)}]{kosterlitz_73}%
  \BibitemOpen
  \bibfield  {author} {\bibinfo {author} {\bibfnamefont {J.~M.}\ \bibnamefont
  {Kosterlitz}}\ and\ \bibinfo {author} {\bibfnamefont {D.~J.}\ \bibnamefont
  {Thouless}},\ }\href {http://stacks.iop.org/0022-3719/6/i=7/a=010} {\bibfield
   {journal} {\bibinfo  {journal} {Journal of Physics C: Solid State Physics}\
  }\textbf {\bibinfo {volume} {6}},\ \bibinfo {pages} {1181} (\bibinfo {year}
  {1973})}\BibitemShut {NoStop}%
\bibitem [{\citenamefont {Kosterlitz}(2016)}]{kosterlitz_16}%
  \BibitemOpen
  \bibfield  {author} {\bibinfo {author} {\bibfnamefont {J.~M.}\ \bibnamefont
  {Kosterlitz}},\ }\href {http://stacks.iop.org/0034-4885/79/i=2/a=026001}
  {\bibfield  {journal} {\bibinfo  {journal} {Rep. Prog. Phys.}\ }\textbf
  {\bibinfo {volume} {79}},\ \bibinfo {pages} {026001} (\bibinfo {year}
  {2016})}\BibitemShut {NoStop}%
\bibitem [{\citenamefont {Ville}\ \emph {et~al.}(2018)\citenamefont {Ville},
  \citenamefont {Saint-Jalm}, \citenamefont {Le~Cerf}, \citenamefont
  {Aidelsburger}, \citenamefont {Nascimb\`ene}, \citenamefont {Dalibard},\ and\
  \citenamefont {Beugnon}}]{ville_18}%
  \BibitemOpen
  \bibfield  {author} {\bibinfo {author} {\bibfnamefont {J.~L.}\ \bibnamefont
  {Ville}}, \bibinfo {author} {\bibfnamefont {R.}~\bibnamefont {Saint-Jalm}},
  \bibinfo {author} {\bibfnamefont {E.}~\bibnamefont {Le~Cerf}}, \bibinfo
  {author} {\bibfnamefont {M.}~\bibnamefont {Aidelsburger}}, \bibinfo {author}
  {\bibfnamefont {S.}~\bibnamefont {Nascimb\`ene}}, \bibinfo {author}
  {\bibfnamefont {J.}~\bibnamefont {Dalibard}}, \ and\ \bibinfo {author}
  {\bibfnamefont {J.}~\bibnamefont {Beugnon}},\ }\href {\doibase
  10.1103/PhysRevLett.121.145301} {\bibfield  {journal} {\bibinfo  {journal}
  {Phys. Rev. Lett.}\ }\textbf {\bibinfo {volume} {121}},\ \bibinfo {pages}
  {145301} (\bibinfo {year} {2018})}\BibitemShut {NoStop}%
\bibitem [{\citenamefont {Suthar}\ \emph {et~al.}(2015)\citenamefont {Suthar},
  \citenamefont {Roy},\ and\ \citenamefont {Angom}}]{suthar_15}%
  \BibitemOpen
  \bibfield  {author} {\bibinfo {author} {\bibfnamefont {K.}~\bibnamefont
  {Suthar}}, \bibinfo {author} {\bibfnamefont {A.}~\bibnamefont {Roy}}, \ and\
  \bibinfo {author} {\bibfnamefont {D.}~\bibnamefont {Angom}},\ }\href
  {\doibase 10.1103/PhysRevA.91.043615} {\bibfield  {journal} {\bibinfo
  {journal} {Phys. Rev. A}\ }\textbf {\bibinfo {volume} {91}},\ \bibinfo
  {pages} {043615} (\bibinfo {year} {2015})}\BibitemShut {NoStop}%
\bibitem [{\citenamefont {Suthar}\ and\ \citenamefont
  {Angom}(2016)}]{suthar_16}%
  \BibitemOpen
  \bibfield  {author} {\bibinfo {author} {\bibfnamefont {K.}~\bibnamefont
  {Suthar}}\ and\ \bibinfo {author} {\bibfnamefont {D.}~\bibnamefont {Angom}},\
  }\href {\doibase 10.1103/PhysRevA.93.063608} {\bibfield  {journal} {\bibinfo
  {journal} {Phys. Rev. A}\ }\textbf {\bibinfo {volume} {93}},\ \bibinfo
  {pages} {063608} (\bibinfo {year} {2016})}\BibitemShut {NoStop}%
\bibitem [{\citenamefont {Bai}\ and\ \citenamefont {Li}(2012)}]{bai_12}%
  \BibitemOpen
  \bibfield  {author} {\bibinfo {author} {\bibfnamefont {Z.}~\bibnamefont
  {Bai}}\ and\ \bibinfo {author} {\bibfnamefont {R.-C.}\ \bibnamefont {Li}},\
  }\href {\doibase 10.1137/110838960} {\bibfield  {journal} {\bibinfo
  {journal} {Siam J. Matrix Anal. Appl .}\ }\textbf {\bibinfo {volume} {33}},\
  \bibinfo {pages} {1075} (\bibinfo {year} {2012})}\BibitemShut {NoStop}%
\bibitem [{\citenamefont {Bai}\ and\ \citenamefont {Li}(2014)}]{bai_14}%
  \BibitemOpen
  \bibfield  {author} {\bibinfo {author} {\bibfnamefont {Z.}~\bibnamefont
  {Bai}}\ and\ \bibinfo {author} {\bibfnamefont {R.-C.}\ \bibnamefont {Li}},\
  }\href {\doibase 10.1007/s10543-014-0472-6} {\bibfield  {journal} {\bibinfo
  {journal} {BIT Numerical Mathematics}\ }\textbf {\bibinfo {volume} {54}},\
  \bibinfo {pages} {31} (\bibinfo {year} {2014})}\BibitemShut {NoStop}%
\bibitem [{\citenamefont {Petrov}\ \emph
  {et~al.}(2000{\natexlab{a}})\citenamefont {Petrov}, \citenamefont
  {Holzmann},\ and\ \citenamefont {Shlyapnikov}}]{petrov_00}%
  \BibitemOpen
  \bibfield  {author} {\bibinfo {author} {\bibfnamefont {D.~S.}\ \bibnamefont
  {Petrov}}, \bibinfo {author} {\bibfnamefont {M.}~\bibnamefont {Holzmann}}, \
  and\ \bibinfo {author} {\bibfnamefont {G.~V.}\ \bibnamefont {Shlyapnikov}},\
  }\href {\doibase 10.1103/PhysRevLett.84.2551} {\bibfield  {journal} {\bibinfo
   {journal} {Phys. Rev. Lett.}\ }\textbf {\bibinfo {volume} {84}},\ \bibinfo
  {pages} {2551} (\bibinfo {year} {2000}{\natexlab{a}})}\BibitemShut {NoStop}%
\bibitem [{\citenamefont {Salasnich}\ \emph {et~al.}(2002)\citenamefont
  {Salasnich}, \citenamefont {Parola},\ and\ \citenamefont
  {Reatto}}]{salasnich_02}%
  \BibitemOpen
  \bibfield  {author} {\bibinfo {author} {\bibfnamefont {L.}~\bibnamefont
  {Salasnich}}, \bibinfo {author} {\bibfnamefont {A.}~\bibnamefont {Parola}}, \
  and\ \bibinfo {author} {\bibfnamefont {L.}~\bibnamefont {Reatto}},\ }\href
  {\doibase 10.1103/PhysRevA.65.043614} {\bibfield  {journal} {\bibinfo
  {journal} {Phys. Rev. A}\ }\textbf {\bibinfo {volume} {65}},\ \bibinfo
  {pages} {043614} (\bibinfo {year} {2002})}\BibitemShut {NoStop}%
\bibitem [{\citenamefont {Petrov}\ \emph
  {et~al.}(2000{\natexlab{b}})\citenamefont {Petrov}, \citenamefont
  {Shlyapnikov},\ and\ \citenamefont {Walraven}}]{petrov_00_1}%
  \BibitemOpen
  \bibfield  {author} {\bibinfo {author} {\bibfnamefont {D.~S.}\ \bibnamefont
  {Petrov}}, \bibinfo {author} {\bibfnamefont {G.~V.}\ \bibnamefont
  {Shlyapnikov}}, \ and\ \bibinfo {author} {\bibfnamefont {J.~T.~M.}\
  \bibnamefont {Walraven}},\ }\href {\doibase 10.1103/PhysRevLett.85.3745}
  {\bibfield  {journal} {\bibinfo  {journal} {Phys. Rev. Lett.}\ }\textbf
  {\bibinfo {volume} {85}},\ \bibinfo {pages} {3745} (\bibinfo {year}
  {2000}{\natexlab{b}})}\BibitemShut {NoStop}%
\bibitem [{\citenamefont {Lehoucq}\ \emph {et~al.}(1998)\citenamefont
  {Lehoucq}, \citenamefont {Sorensen},\ and\ \citenamefont
  {Yang}}]{lehoucq_98}%
  \BibitemOpen
  \bibfield  {author} {\bibinfo {author} {\bibfnamefont {R.}~\bibnamefont
  {Lehoucq}}, \bibinfo {author} {\bibfnamefont {D.}~\bibnamefont {Sorensen}}, \
  and\ \bibinfo {author} {\bibfnamefont {C.}~\bibnamefont {Yang}},\ }\href
  {\doibase 10.1137/1.9780898719628} {\emph {\bibinfo {title} {ARPACK Users'
  Guide}}}\ (\bibinfo  {publisher} {Society for Industrial and Applied
  Mathematics (Philadelphia)},\ \bibinfo {year} {1998})\BibitemShut {NoStop}%
\bibitem [{\citenamefont {Jain}\ and\ \citenamefont
  {Boninsegni}(2011)}]{jain_11}%
  \BibitemOpen
  \bibfield  {author} {\bibinfo {author} {\bibfnamefont {P.}~\bibnamefont
  {Jain}}\ and\ \bibinfo {author} {\bibfnamefont {M.}~\bibnamefont
  {Boninsegni}},\ }\href {\doibase 10.1103/PhysRevA.83.023602} {\bibfield
  {journal} {\bibinfo  {journal} {Phys. Rev. A}\ }\textbf {\bibinfo {volume}
  {83}},\ \bibinfo {pages} {023602} (\bibinfo {year} {2011})}\BibitemShut
  {NoStop}%
\bibitem [{\citenamefont {Ticknor}(2014)}]{ticknor_14}%
  \BibitemOpen
  \bibfield  {author} {\bibinfo {author} {\bibfnamefont {C.}~\bibnamefont
  {Ticknor}},\ }\href {\doibase 10.1103/PhysRevA.89.053601} {\bibfield
  {journal} {\bibinfo  {journal} {Phys. Rev. A}\ }\textbf {\bibinfo {volume}
  {89}},\ \bibinfo {pages} {053601} (\bibinfo {year} {2014})}\BibitemShut
  {NoStop}%
\bibitem [{\citenamefont {Wilson}\ \emph {et~al.}(2010)\citenamefont {Wilson},
  \citenamefont {Ronen},\ and\ \citenamefont {Bohn}}]{wilson_10}%
  \BibitemOpen
  \bibfield  {author} {\bibinfo {author} {\bibfnamefont {R.~M.}\ \bibnamefont
  {Wilson}}, \bibinfo {author} {\bibfnamefont {S.}~\bibnamefont {Ronen}}, \
  and\ \bibinfo {author} {\bibfnamefont {J.~L.}\ \bibnamefont {Bohn}},\ }\href
  {\doibase 10.1103/PhysRevLett.104.094501} {\bibfield  {journal} {\bibinfo
  {journal} {Phys. Rev. Lett.}\ }\textbf {\bibinfo {volume} {104}},\ \bibinfo
  {pages} {094501} (\bibinfo {year} {2010})}\BibitemShut {NoStop}%
\bibitem [{\citenamefont {Zambelli}\ \emph {et~al.}(2000)\citenamefont
  {Zambelli}, \citenamefont {Pitaevskii}, \citenamefont {Stamper-Kurn},\ and\
  \citenamefont {Stringari}}]{zambelli_2000}%
  \BibitemOpen
  \bibfield  {author} {\bibinfo {author} {\bibfnamefont {F.}~\bibnamefont
  {Zambelli}}, \bibinfo {author} {\bibfnamefont {L.}~\bibnamefont
  {Pitaevskii}}, \bibinfo {author} {\bibfnamefont {D.~M.}\ \bibnamefont
  {Stamper-Kurn}}, \ and\ \bibinfo {author} {\bibfnamefont {S.}~\bibnamefont
  {Stringari}},\ }\href {\doibase 10.1103/PhysRevA.61.063608} {\bibfield
  {journal} {\bibinfo  {journal} {Phys. Rev. A}\ }\textbf {\bibinfo {volume}
  {61}},\ \bibinfo {pages} {063608} (\bibinfo {year} {2000})}\BibitemShut
  {NoStop}%
\bibitem [{\citenamefont {Stamper-Kurn}\ \emph {et~al.}(1999)\citenamefont
  {Stamper-Kurn}, \citenamefont {Chikkatur}, \citenamefont {G\"orlitz},
  \citenamefont {Inouye}, \citenamefont {Gupta}, \citenamefont {Pritchard},\
  and\ \citenamefont {Ketterle}}]{stamper-kurn_99}%
  \BibitemOpen
  \bibfield  {author} {\bibinfo {author} {\bibfnamefont {D.~M.}\ \bibnamefont
  {Stamper-Kurn}}, \bibinfo {author} {\bibfnamefont {A.~P.}\ \bibnamefont
  {Chikkatur}}, \bibinfo {author} {\bibfnamefont {A.}~\bibnamefont
  {G\"orlitz}}, \bibinfo {author} {\bibfnamefont {S.}~\bibnamefont {Inouye}},
  \bibinfo {author} {\bibfnamefont {S.}~\bibnamefont {Gupta}}, \bibinfo
  {author} {\bibfnamefont {D.~E.}\ \bibnamefont {Pritchard}}, \ and\ \bibinfo
  {author} {\bibfnamefont {W.}~\bibnamefont {Ketterle}},\ }\href {\doibase
  10.1103/PhysRevLett.83.2876} {\bibfield  {journal} {\bibinfo  {journal}
  {Phys. Rev. Lett.}\ }\textbf {\bibinfo {volume} {83}},\ \bibinfo {pages}
  {2876} (\bibinfo {year} {1999})}\BibitemShut {NoStop}%
\bibitem [{\citenamefont {Stenger}\ \emph {et~al.}(1999)\citenamefont
  {Stenger}, \citenamefont {Inouye}, \citenamefont {Chikkatur}, \citenamefont
  {Stamper-Kurn}, \citenamefont {Pritchard},\ and\ \citenamefont
  {Ketterle}}]{stenger_99}%
  \BibitemOpen
  \bibfield  {author} {\bibinfo {author} {\bibfnamefont {J.}~\bibnamefont
  {Stenger}}, \bibinfo {author} {\bibfnamefont {S.}~\bibnamefont {Inouye}},
  \bibinfo {author} {\bibfnamefont {A.~P.}\ \bibnamefont {Chikkatur}}, \bibinfo
  {author} {\bibfnamefont {D.~M.}\ \bibnamefont {Stamper-Kurn}}, \bibinfo
  {author} {\bibfnamefont {D.~E.}\ \bibnamefont {Pritchard}}, \ and\ \bibinfo
  {author} {\bibfnamefont {W.}~\bibnamefont {Ketterle}},\ }\href {\doibase
  10.1103/PhysRevLett.82.4569} {\bibfield  {journal} {\bibinfo  {journal}
  {Phys. Rev. Lett.}\ }\textbf {\bibinfo {volume} {82}},\ \bibinfo {pages}
  {4569} (\bibinfo {year} {1999})}\BibitemShut {NoStop}%
\bibitem [{\citenamefont {Wu}\ and\ \citenamefont {Griffin}(1996)}]{wu_96}%
  \BibitemOpen
  \bibfield  {author} {\bibinfo {author} {\bibfnamefont {W.-C.}\ \bibnamefont
  {Wu}}\ and\ \bibinfo {author} {\bibfnamefont {A.}~\bibnamefont {Griffin}},\
  }\href {\doibase 10.1103/PhysRevA.54.4204} {\bibfield  {journal} {\bibinfo
  {journal} {Phys. Rev. A}\ }\textbf {\bibinfo {volume} {54}},\ \bibinfo
  {pages} {4204} (\bibinfo {year} {1996})}\BibitemShut {NoStop}%
\bibitem [{\citenamefont {Menotti}\ \emph {et~al.}(2003)\citenamefont
  {Menotti}, \citenamefont {Kr\"amer}, \citenamefont {Pitaevskii},\ and\
  \citenamefont {Stringari}}]{menotti_03}%
  \BibitemOpen
  \bibfield  {author} {\bibinfo {author} {\bibfnamefont {C.}~\bibnamefont
  {Menotti}}, \bibinfo {author} {\bibfnamefont {M.}~\bibnamefont {Kr\"amer}},
  \bibinfo {author} {\bibfnamefont {L.}~\bibnamefont {Pitaevskii}}, \ and\
  \bibinfo {author} {\bibfnamefont {S.}~\bibnamefont {Stringari}},\ }\href
  {\doibase 10.1103/PhysRevA.67.053609} {\bibfield  {journal} {\bibinfo
  {journal} {Phys. Rev. A}\ }\textbf {\bibinfo {volume} {67}},\ \bibinfo
  {pages} {053609} (\bibinfo {year} {2003})}\BibitemShut {NoStop}%
\bibitem [{\citenamefont {Muruganandam}\ and\ \citenamefont
  {Adhikari}(2009)}]{muruganandam_09}%
  \BibitemOpen
  \bibfield  {author} {\bibinfo {author} {\bibfnamefont {P.}~\bibnamefont
  {Muruganandam}}\ and\ \bibinfo {author} {\bibfnamefont {S.}~\bibnamefont
  {Adhikari}},\ }\href {\doibase https://doi.org/10.1016/j.cpc.2009.04.015}
  {\bibfield  {journal} {\bibinfo  {journal} {Comput. Phys. Commun.}\ }\textbf
  {\bibinfo {volume} {180}},\ \bibinfo {pages} {1888 } (\bibinfo {year}
  {2009})}\BibitemShut {NoStop}%
\bibitem [{\citenamefont {Vudragovi{\'c}}\ \emph {et~al.}(2012)\citenamefont
  {Vudragovi{\'c}}, \citenamefont {Vidanovi{\'c}}, \citenamefont {Bala{\v z}},
  \citenamefont {Muruganandam},\ and\ \citenamefont
  {Adhikari}}]{vudragovic_12}%
  \BibitemOpen
  \bibfield  {author} {\bibinfo {author} {\bibfnamefont {D.}~\bibnamefont
  {Vudragovi{\'c}}}, \bibinfo {author} {\bibfnamefont {I.}~\bibnamefont
  {Vidanovi{\'c}}}, \bibinfo {author} {\bibfnamefont {A.}~\bibnamefont {Bala{\v
  z}}}, \bibinfo {author} {\bibfnamefont {P.}~\bibnamefont {Muruganandam}}, \
  and\ \bibinfo {author} {\bibfnamefont {S.~K.}\ \bibnamefont {Adhikari}},\
  }\href {\doibase http://dx.doi.org/10.1016/j.cpc.2012.03.022} {\bibfield
  {journal} {\bibinfo  {journal} {Comput. Phys. Commun.}\ }\textbf {\bibinfo
  {volume} {183}},\ \bibinfo {pages} {2021} (\bibinfo {year}
  {2012})}\BibitemShut {NoStop}%
\bibitem [{\citenamefont {Young-S.}\ \emph {et~al.}(2016)\citenamefont
  {Young-S.}, \citenamefont {Vudragovi{\'c}}, \citenamefont {Muruganandam},
  \citenamefont {Adhikari},\ and\ \citenamefont {Bala{\v z}}}]{young_16}%
  \BibitemOpen
  \bibfield  {author} {\bibinfo {author} {\bibfnamefont {L.~E.}\ \bibnamefont
  {Young-S.}}, \bibinfo {author} {\bibfnamefont {D.}~\bibnamefont
  {Vudragovi{\'c}}}, \bibinfo {author} {\bibfnamefont {P.}~\bibnamefont
  {Muruganandam}}, \bibinfo {author} {\bibfnamefont {S.~K.}\ \bibnamefont
  {Adhikari}}, \ and\ \bibinfo {author} {\bibfnamefont {A.}~\bibnamefont
  {Bala{\v z}}},\ }\href {\doibase https://doi.org/10.1016/j.cpc.2016.03.015}
  {\bibfield  {journal} {\bibinfo  {journal} {Comput. Phys. Commun.}\ }\textbf
  {\bibinfo {volume} {204}},\ \bibinfo {pages} {209 } (\bibinfo {year}
  {2016})}\BibitemShut {NoStop}%
\bibitem [{\citenamefont {Young-S.}\ \emph {et~al.}(2017)\citenamefont
  {Young-S.}, \citenamefont {Muruganandam}, \citenamefont {Adhikari},
  \citenamefont {Lon{\v c}ar}, \citenamefont {Vudragovi{\'c}},\ and\
  \citenamefont {Bala{\v z}}}]{young_17}%
  \BibitemOpen
  \bibfield  {author} {\bibinfo {author} {\bibfnamefont {L.~E.}\ \bibnamefont
  {Young-S.}}, \bibinfo {author} {\bibfnamefont {P.}~\bibnamefont
  {Muruganandam}}, \bibinfo {author} {\bibfnamefont {S.~K.}\ \bibnamefont
  {Adhikari}}, \bibinfo {author} {\bibfnamefont {V.}~\bibnamefont {Lon{\v
  c}ar}}, \bibinfo {author} {\bibfnamefont {D.}~\bibnamefont {Vudragovi{\'c}}},
  \ and\ \bibinfo {author} {\bibfnamefont {A.}~\bibnamefont {Bala{\v z}}},\
  }\href {\doibase https://doi.org/10.1016/j.cpc.2017.07.013} {\bibfield
  {journal} {\bibinfo  {journal} {Comput. Phys. Commun.}\ }\textbf {\bibinfo
  {volume} {220}},\ \bibinfo {pages} {503 } (\bibinfo {year}
  {2017})}\BibitemShut {NoStop}%
\bibitem [{\citenamefont {Minguzzi}\ \emph {et~al.}(2004)\citenamefont
  {Minguzzi}, \citenamefont {Succi}, \citenamefont {Toschi}, \citenamefont
  {Tosi},\ and\ \citenamefont {Vignolo}}]{minguzzi_04}%
  \BibitemOpen
  \bibfield  {author} {\bibinfo {author} {\bibfnamefont {A.}~\bibnamefont
  {Minguzzi}}, \bibinfo {author} {\bibfnamefont {S.}~\bibnamefont {Succi}},
  \bibinfo {author} {\bibfnamefont {F.}~\bibnamefont {Toschi}}, \bibinfo
  {author} {\bibfnamefont {M.}~\bibnamefont {Tosi}}, \ and\ \bibinfo {author}
  {\bibfnamefont {P.}~\bibnamefont {Vignolo}},\ }\href {\doibase
  https://doi.org/10.1016/j.physrep.2004.02.001} {\bibfield  {journal}
  {\bibinfo  {journal} {Physics Reports}\ }\textbf {\bibinfo {volume} {395}},\
  \bibinfo {pages} {223 } (\bibinfo {year} {2004})}\BibitemShut {NoStop}%
\bibitem [{\citenamefont {Antoine}\ \emph {et~al.}(2013)\citenamefont
  {Antoine}, \citenamefont {Bao},\ and\ \citenamefont {Besse}}]{antoine_13}%
  \BibitemOpen
  \bibfield  {author} {\bibinfo {author} {\bibfnamefont {X.}~\bibnamefont
  {Antoine}}, \bibinfo {author} {\bibfnamefont {W.}~\bibnamefont {Bao}}, \ and\
  \bibinfo {author} {\bibfnamefont {C.}~\bibnamefont {Besse}},\ }\href
  {\doibase https://doi.org/10.1016/j.cpc.2013.07.012} {\bibfield  {journal}
  {\bibinfo  {journal} {Computer Physics Communications}\ }\textbf {\bibinfo
  {volume} {184}},\ \bibinfo {pages} {2621} (\bibinfo {year}
  {2013})}\BibitemShut {NoStop}%
\bibitem [{\citenamefont {Bao}(2004)}]{bao_04}%
  \BibitemOpen
  \bibfield  {author} {\bibinfo {author} {\bibfnamefont {W.}~\bibnamefont
  {Bao}},\ }\href@noop {} {\bibfield  {journal} {\bibinfo  {journal}
  {Multiscale Model. Simul.}\ }\textbf {\bibinfo {volume} {2}},\ \bibinfo
  {pages} {210} (\bibinfo {year} {2004})}\BibitemShut {NoStop}%
\bibitem [{\citenamefont {Bao}\ and\ \citenamefont {Shen}(2005)}]{bao_05}%
  \BibitemOpen
  \bibfield  {author} {\bibinfo {author} {\bibfnamefont {W.}~\bibnamefont
  {Bao}}\ and\ \bibinfo {author} {\bibfnamefont {J.}~\bibnamefont {Shen}},\
  }\href@noop {} {\bibfield  {journal} {\bibinfo  {journal} {SIAM J. Sci.
  Comput.}\ }\textbf {\bibinfo {volume} {26}},\ \bibinfo {pages} {2010}
  (\bibinfo {year} {2005})}\BibitemShut {NoStop}%
\bibitem [{\citenamefont {Tiwari}\ and\ \citenamefont
  {Shukla}(2006)}]{tiwari_06}%
  \BibitemOpen
  \bibfield  {author} {\bibinfo {author} {\bibfnamefont {R.~P.}\ \bibnamefont
  {Tiwari}}\ and\ \bibinfo {author} {\bibfnamefont {A.}~\bibnamefont
  {Shukla}},\ }\href {\doibase https://doi.org/10.1016/j.cpc.2005.10.014}
  {\bibfield  {journal} {\bibinfo  {journal} {Computer Physics Communications}\
  }\textbf {\bibinfo {volume} {174}},\ \bibinfo {pages} {966} (\bibinfo {year}
  {2006})}\BibitemShut {NoStop}%
\bibitem [{\citenamefont {Simula}\ \emph {et~al.}(2001)\citenamefont {Simula},
  \citenamefont {Virtanen},\ and\ \citenamefont {Salomaa}}]{simula_01}%
  \BibitemOpen
  \bibfield  {author} {\bibinfo {author} {\bibfnamefont {T.}~\bibnamefont
  {Simula}}, \bibinfo {author} {\bibfnamefont {S.}~\bibnamefont {Virtanen}}, \
  and\ \bibinfo {author} {\bibfnamefont {M.}~\bibnamefont {Salomaa}},\ }\href
  {\doibase http://dx.doi.org/10.1016/S0010-4655(01)00369-1} {\bibfield
  {journal} {\bibinfo  {journal} {Comput. Phys. Commun.}\ }\textbf {\bibinfo
  {volume} {142}},\ \bibinfo {pages} {396} (\bibinfo {year}
  {2001})}\BibitemShut {NoStop}%
\end{thebibliography}%

\section{Appendix}
The explicit forms of $\hat{H}_n^i$, where $0\leqslant n\leqslant 4$ and 
$i=1,2$ represent the order of fluctuations and species index, are
\begin{eqnarray}
 \hat{H}_0^1 &=& \iint dxdy\,
 \phi_1^*\left(\hat{h}_1-\mu_1+\frac{U_{11}}{2}|\phi_1|^2
  +\frac{U_{12}}{2}|\phi_2|^2\right)\phi_1,\nonumber\\
  \hat{H}_0^2 &=& \iint dxdy\,
 \phi_2^*\left(\hat{h}_2-\mu_2+\frac{U_{22}}{2}|\phi_2|^2
  +\frac{U_{12}}{2}|\phi_1|^2\right)\phi_2,\nonumber\\
 \hat{H}_1^1 &=& \iint dxdy\, \left[\phi_1^*\left(\hat{h}_1-\mu_1 + U_{11}|
 \phi_1|^2
  + U_{12}|\phi_2|^2\right)\tilde{\psi}_1
 + \tilde{\psi}_1^\dagger\left(\hat{h}_1-\mu_1+U_{11}|\phi_1|^2
  + U_{12}|\phi_2|^2\right)\phi_1\right],\nonumber\\  
 \hat{H}_1^2 &=& \iint dxdy\, \left[\phi_2^*\left(\hat{h}_2-\mu_2 
 + U_{22}|\phi_2|^2
  + U_{12}|\phi_1|^2\right)\tilde{\psi}_2
 + \tilde{\psi}_2^\dagger\left(\hat{h}_2-\mu_2+U_{22}|\phi_2|^2
  + U_{12}|\phi_1|^2\right)\phi_2\right],\nonumber\\
\hat{H}_2^1 &=&\iint dxdy\, \bigg[\tilde{\psi_1}^\dagger\left(\hat{h}_1-\mu_1
+2U_{11}
 |\phi_1|^2 +U_{12}|\phi_2|^2\right)\tilde{\psi}_1+
 \frac{U_{11}}{2}\left(\phi_1^{*2}\tilde{\psi}_1\tilde{\psi}_1
 + \phi_1^2\tilde{\psi}_1^\dagger\tilde{\psi}_1^\dagger\right)\nonumber\\
&& +\frac{U_{12}}{2}\left(\phi_1^*\phi_2^*\tilde{\psi}_1\tilde{\psi}_2
 +\phi_1^*\phi_2\tilde{\psi}_2^\dagger\tilde{\psi}_1
 +\phi_1\phi_2^*\tilde{\psi}_1^\dagger\tilde{\psi}_2
 +\phi_1\phi_2\tilde{\psi}_1^\dagger\tilde{\psi}_2^\dagger\right)\bigg],
 \nonumber\nonumber\\
\hat{H}_2^2 &=&\iint dxdy\, \bigg[\tilde{\psi_2}^\dagger\left(\hat{h}_2-\mu_2
+2U_{22}
 |\phi_2|^2 +U_{12}|\phi_1|^2\right)\tilde{\psi}_2+
 \frac{U_{22}}{2}\left(\phi_2^{*2}\tilde{\psi}_2\tilde{\psi}_2
 + \phi_2^2\tilde{\psi}_2^\dagger\tilde{\psi}_2^\dagger\right)\nonumber\\
&&+\frac{U_{12}}{2}\left(\phi_1^*\phi_2^*\tilde{\psi}_1\tilde{\psi}_2
 +\phi_1^*\phi_2\tilde{\psi}_2^\dagger\tilde{\psi}_1
 +\phi_1\phi_2^*\tilde{\psi}_1^\dagger\tilde{\psi}_2
 +\phi_1\phi_2\tilde{\psi}_1^\dagger\tilde{\psi}_2^\dagger\right)\bigg],
 \nonumber\\
 \hat{H}_3^1 &=& \iint dxdy\,\bigg[
 U_{11}\left(\phi_1^*\tilde{\psi}_1^\dagger\tilde{\psi}_1\tilde{\psi}_1
 +\phi_1\tilde{\psi}_1^\dagger\tilde{\psi}_1^\dagger\tilde{\psi}_1\right)
 +\frac{U_{12}}{2}\big(\phi_1^*\tilde{\psi}_2^\dagger\tilde{\psi}_1
\tilde{\psi}_2 + \phi_2^*\tilde{\psi}_1^\dagger\tilde{\psi}_1\tilde{\psi}_2
\nonumber\\
&&+\phi_1\tilde{\psi}_1^\dagger\tilde{\psi}_2^\dagger\tilde{\psi}_2
+\phi_2\tilde{\psi}_1^\dagger\tilde{\psi}_2^\dagger\tilde{\psi}_1\big)\bigg],
\nonumber
\end{eqnarray}
\begin{eqnarray}
\hat{H}_3^2 &=& \iint dxdy\,\bigg[
 U_{22}\left(\phi_2^*\tilde{\psi}_2^\dagger\tilde{\psi}_2\tilde{\psi}_2
 +\phi_2\tilde{\psi}_2^\dagger\tilde{\psi}_2^\dagger\tilde{\psi}_2\right)
 +\frac{U_{12}}{2}\big(\phi_1^*\tilde{\psi}_2^\dagger\tilde{\psi}_1
\tilde{\psi}_2 + \phi_2^*\tilde{\psi}_1^\dagger\tilde{\psi}_1\tilde{\psi}_2
\nonumber\\
&&+\phi_1\tilde{\psi}_1^\dagger\tilde{\psi}_2^\dagger\tilde{\psi}_2
+\phi_2\tilde{\psi}_1^\dagger\tilde{\psi}_2^\dagger\tilde{\psi}_1\big)\bigg],
\nonumber\\
\hat{H}_4^1 &=&\iint dxdy\, \bigg[\frac{U_{11}}{2}\tilde{\psi}_1^\dagger
\tilde{\psi}_1
^\dagger\tilde{\psi}_1\tilde{\psi}_1 + \frac{U_{12}}{2}\tilde{\psi}_1^\dagger
\tilde{\psi}_2^\dagger\tilde{\psi}_1\tilde{\psi}_2\bigg],\nonumber\\
\hat{H}_4^2 &=&\iint dxdy\, \bigg[\frac{U_{22}}{2}\tilde{\psi}_2^\dagger
\tilde{\psi}_2
^\dagger\tilde{\psi}_2\tilde{\psi}_2 + \frac{U_{12}}{2}\tilde{\psi}_1^\dagger
\tilde{\psi}_2^\dagger\tilde{\psi}_1\tilde{\psi}_2\bigg].
\label{hfbham}
\end{eqnarray}

Using the definition of field operator from Eq.~(\ref{ch2dectbec}) and
putting it in
Eq.~(\ref{twocomp}), the Heisenberg equation of motion for the first 
species ( $i=1$) is
\begin{eqnarray}
  i\hbar\frac{\partial(\phi_{1}+\tilde{\psi}_1)}{\partial t}
    &=&\left[\frac{-\hbar^2}{2m_1}\nabla^2 \phi_{1}  
    -\frac{\hbar^2}{2m_1}\nabla^2 \tilde{\psi_{1}} + V_1\phi_1 
    + V_1\tilde{\psi}_1 \right.      \nonumber\\
    && \left. +U_{11}\hat{\Psi}_{1}^{\dagger}\hat{\Psi}_{1}\hat{\Psi}_{1}
    +U_{12}\hat{\Psi}_{2}^{\dagger}\hat{\Psi}_{2}\hat{\Psi}_{1}-\mu_{1}\phi_1
    -\mu_{1}\tilde{\psi}_1\right].
\label{eq2}
\end{eqnarray}
The interaction terms in the equation can be written in terms of $c$-number
and fluctuation operators as
\begin{subequations}
  \begin{eqnarray}
     \hat{\Psi}_{1}^{\dagger}\hat{\Psi}_{1}\hat{\Psi}_{1}
        &=&|\phi_1|^{2}\phi_1 + 2|\phi_1|^{2}\tilde{\psi}_1+2 \phi_{1}
        \tilde{\psi}^{\dagger}_{1}\tilde{\psi}_{1} 
        +\phi_{1}^{*}\tilde{\psi}_{1}\tilde{\psi}_{1}
        +\phi_{1}^{2}\tilde{\psi}_1^{\dagger}
        +\tilde{\psi}_{1}^{\dagger}\tilde{\psi}_1\tilde{\psi}_1,\;\;\;\;\;\;\\
    \hat{\Psi}_{2}^{\dagger}\hat{\Psi}_{2}\hat{\Psi}_{1}
        &=&|\phi_2|^{2}\phi_1+ |\phi_2|^{2}\tilde{\psi}_1+\phi_{2}^{*}
        \tilde{\psi}_{2}\phi_{1}+\phi_{2}^{*}\tilde{\psi}_{2}\tilde{\psi}_{1}
        +\tilde{\psi}_{2}^{\dagger}\phi_{2}\phi_{1} 
        +\tilde{\psi}_{2}^{\dagger}\phi_{2}\tilde{\psi}_{1}    \nonumber  \\
        &&+\tilde{\psi}_2^{\dagger}\tilde{\psi}_2\phi_1 
        +\tilde{\psi}_2^{\dagger}\tilde{\psi}_2\tilde{\psi}_1.
    \label{int2}
  \end{eqnarray}
\end{subequations}
Since all the atomic fluctuations (quantum and thermal) associated in this
theory are white noise
$\langle\tilde{\psi_i}\rangle=\langle\tilde{\psi_i}^{\dagger}\rangle=0$.
Hence the expectation value of the product of operators are
\begin{subequations}
  \begin{eqnarray}
     \langle\hat{\Psi}_{1}^{\dagger}\hat{\Psi}_{1}\hat{\Psi}_{1}\rangle
        &=&|\phi_1|^{2}\phi_1+\phi_{1}^{*}\langle\tilde{\psi}_{1}
        \tilde{\psi}_{1}\rangle+ 2\phi_{1}\langle\tilde{\psi}^{\dagger}_{1}
        \tilde{\psi}_{1}\rangle +\langle\tilde{\psi}_1^{\dagger}
        \tilde{\psi}_1\tilde{\psi}_1\rangle,\;\;\;\;\;\; \\
     \langle\hat{\Psi}_{2}^{\dagger}\hat{\Psi}_{2}\hat{\Psi}_{1}\rangle
        &=& |\phi_1|^{2}\phi_1+\phi_{2}^{*}\langle\tilde{\psi}_{2}
        \tilde{\psi}_{1}\rangle +\phi_2\langle\tilde{\psi}_{2}^{\dagger}
        \tilde{\psi}_{1}\rangle +\phi_1\langle\tilde{\psi}_2^{\dagger}
        \tilde{\psi}_2\rangle             \nonumber\\
        &&+\langle\tilde{\psi}_2^{\dagger}\tilde{\psi}_2\tilde{\psi}_1\rangle.
  \end{eqnarray}
\end{subequations}
Considering that the fluctuations of the two species are uncorrelated
$\langle\tilde{\psi}_{2}\tilde{\psi}_{1}\rangle=\langle
\tilde{\psi}_{2}^{\dagger}\tilde{\psi}_{1}\rangle=0$, the equation of motion 
of the condensate of the first species is obtained by
taking the average of Eq.~(\ref{eq2}) as
\begin{eqnarray}
   i\hbar\frac{\partial \phi_1}{\partial t} 
     &=& \left[-\frac{\hbar^{2}}{2m_1}\nabla^2+V_1-\mu_1\right]\phi_1 
     + U_{11}\left[n_{1c}+2\tilde{n}_{1}\right]\phi_1
     +U_{11}\tilde{m}_1\phi_{1}^{*}\nonumber\\
     &&+U_{12}\left[n_{2c}+\tilde{n}_2\right]\phi_1
     +\langle\tilde{\psi}_1^{\dagger} \tilde{\psi}_1\tilde{\psi}_1\rangle
     +\langle\tilde{\psi}_2^{\dagger}\tilde{\psi}_2\tilde{\psi}_1\rangle.
  \label{condeq1}
\end{eqnarray}
Similarly, the equation of motion for the condensate of the second species 
is 
\begin{eqnarray}
  i\hbar\frac{\partial \phi_2}{\partial t}
    &=&\left[-\frac{\hbar^{2}}{2m_2}\nabla^2+V_2-\mu_2\right]\phi_2 
    + U_{22}\left[n_{2c}+2\tilde{n}_{2}\right]\phi_2
    +U_{22}\tilde{m}_2\phi_{2}^{*}     \nonumber\\
    &&+U_{12}\left[n_{1c}+\tilde{n}_1\right]\phi_2
    +\langle\tilde{\psi}_2^{\dagger}\tilde{\psi}_2\tilde{\psi}_2\rangle
    +\langle\tilde{\psi}_1^\dagger\tilde{\psi}_1\tilde{\psi_2}\rangle,
\end{eqnarray}
where we have introduced the local densities:
$n_{ic}\equiv|\phi_i|^2$, $\tilde{n}_i\equiv\langle\tilde{\psi}_{i}
^{\dagger}\tilde{\psi}_i\rangle$, $\tilde{m}_i\equiv\langle\tilde{\psi}_{i}
\tilde{\psi}_i\rangle$
as the condensate, non-condensate, and anomalous densities, respectively.
The equation of motion for the 
non-condensate density of the first species is 
\begin{equation}
  i\hbar\frac{\partial \tilde{\psi}_1}{\partial
  t}=i\hbar\frac{\partial}{\partial t}(\hat{\psi}_1-\phi_1).
\end{equation}
Using Eq.~(\ref{eq2}) and Eq.~(\ref{condeq1}) and applying mean-field 
approximation,
$\tilde{\psi}_i^\dagger\tilde{\psi}_j \simeq\langle\tilde{\psi}_i^\dagger
\tilde{\psi}_j\rangle$, 
$\tilde{\psi}_i \tilde{\psi}_j \simeq
\langle\tilde{\psi}_i\tilde{\psi}_j\rangle$, 
$\tilde{\psi}_1^\dagger\tilde{\psi}_1\tilde{\psi}_1\simeq2
\langle\tilde{\psi}_1^\dagger\tilde{\psi}_1\rangle\tilde{\psi}_1
+\langle\tilde{\psi}_1\tilde{\psi}_1\rangle\tilde{\psi}_1^\dagger$,
$\tilde{\psi}_2^\dagger\tilde{\psi}_2\tilde{\psi}_1\simeq\langle$
$\tilde{\psi}_2^\dagger\tilde{\psi}_2\rangle\tilde{\psi}_1$, we can derive the 
equation of motion of the fluctuation operators.

\end{document}